\documentclass[10pt,
							 twocolumn,
							 superscriptaddress,
							 english,
							 pra,
							 showpacs,
							 floatfix,
							 aps
							]{revtex4-2}
\usepackage[utf8]{inputenc}
\usepackage{amsmath}
\usepackage{amssymb}
\usepackage{graphicx}
\usepackage{xspace}
\usepackage{mathtools}
\usepackage{comment}
\usepackage{physics}
\usepackage{bm}

\makeatletter
 
\usepackage[normalem]{ulem}
\usepackage{xcolor}
\usepackage{soul}
\usepackage[backref=none,
bookmarksnumbered=true,
bookmarks=true,
bookmarksopen=true,
colorlinks=true,
citecolor=blue,
linkcolor=blue,
anchorcolor=green,
urlcolor=blue,unicode=false]{hyperref}

\makeatother

\usepackage{babel}

\begin{document}

\title{Quantum Yu-Shiba-Rusinov dimers}

\author{Harald Schmid}
\affiliation{\mbox{Dahlem Center for Complex Quantum Systems and Fachbereich Physik, Freie Universit\"at Berlin, 14195 Berlin, Germany}}

\author{Jacob F.\ Steiner}
\affiliation{\mbox{Dahlem Center for Complex Quantum Systems and Fachbereich Physik, Freie Universit\"at Berlin, 14195 Berlin, Germany}}

\author{Katharina J. Franke}
\affiliation{\mbox{Fachbereich Physik, Freie Universit\"at Berlin, 14195 Berlin, Germany}}

\author{Felix von Oppen}
\affiliation{\mbox{Dahlem Center for Complex Quantum Systems and Fachbereich Physik, Freie Universit\"at Berlin, 14195 Berlin, Germany}}

\date{\today}

\begin{abstract}
Magnetic adatoms on a superconducting substrate undergo a quantum phase transition as their exchange coupling to the conduction electrons increases. For quantum spins, this transition is accompanied by screening of the adatom spin. Here, we explore the consequences of this screening for the phase diagrams and subgap excitation spectra of dimers of magnetic adatoms coupled by hybridization of their Yu-Shiba-Rusinov states and spin-spin interactions. We specifically account for higher spins, single-ion anisotropy, Ruderman-Kittel-Kasuya-Yosida coupling, and Dzyaloshinsky-Moriya interactions relevant in transition-metal and rare-earth systems. Our flexible approach based on a zero-bandwidth approximation provides detailed physical insight and is in excellent qualitative agreement with available numerical-renormalization group calculations on monomers and dimers. Remarkably, we find that even in the limit of large impurity spins or strong single-ion anisotropy, the phase diagrams for dimers of quantum spins remain qualitatively distinct from phase diagrams based on classical spins, highlighting the need for a theory of quantum Yu-Shiba-Rusinov dimers. 
\end{abstract}

\pacs{%
			} 
\maketitle 

\section{Introduction}

Assemblies of magnetic adatoms on superconductors are currently attracting much attention as platforms for topological superconductivity \cite{Pawlak2019,Jack2021,Flensberg2021} and correlated electron physics \cite{Steiner2021}. The adatoms induce Yu-Shiba-Rusinov (YSR) states within the excitation gap of the substrate superconductor \cite{Yu1965,Shiba1968,Rusinov1969,Yazdani1997,Ji2008, Franke2011,Balatsky2006, Heinrich2018}, which hybridize between adjacent sites of the assembly. Several recent experiments \cite{Ruby2018, Choi2018, Kezilebieke2018, Beck2020, Ding2021, Kamlapure2018, Kuester2021,Liebhaber2021, Huang2020} probe adatom dimers, which constitute the minimal example of such an assembly. If the adatoms of the dimer are spaced such that their atomic $d$ orbitals do not overlap, the coupling is entirely mediated by the superconducting substrate. The familiar Ruderman-Kittel-Kasuya-Yosida (RKKY) \cite{Ruderman1954, Kasuya1956, Yosida1957,Anderson1959} and Dzyaloshinsky-Moriya (DM) \cite{Dzyaloshinsky1958,Moriya1960} interactions between the adatom spins are complemented by hybridization of their YSR states. 

So far, dimer experiments have been largely interpreted assuming classical impurity spins. In this framework, the impurity spin acts as a local Zeeman field on the substrate superconductor and the coupling in the dimer depends on the relative orientation of the adatom spins \cite{Rusinov1969,Pientka2013,Poyhonen2014,Meng2015,Hoffman2015,Brydon2015,Ruby2018,Korber2018,Ding2021}. In the absence of spin-orbit coupling, the YSR states split for ferromagnetic alignment, but remain unsplit for antiferromagnetic spins. There is also an overall shift of the YSR levels of the dimer relative to the monomer, which tends to be small compared to the splitting for the ferromagnetic dimer \cite{Rusinov1969,Liebhaber2021}. 

Experiments on individual adatoms on superconductors suggest, however, that their spins are quantum. In particular, this is implied by the observation of Kondo resonances, both on normal-metal \cite{Madhavan1998,Li1998} and superconducting substrates \cite{Franke2011,Hatter2017,Kamlapure2018,Farinacci2020, Verdu2021, Kamlapure2019,Odobesko2020}, and of discrete spin excitations in the presence of single-ion anisotropy \cite{Hirjibehedin2007,Tsukuhara2009, Heinrich2013b, Kezilebieke2019}. Dimers of quantum spins on a superconducting substrate were discussed by Zitko et al.\ \cite{Zitko2011} and 
Yao et al.\ \cite{Yao2014}, based on the numerical renormalization group (NRG). While these calculations were limited to spin-$\frac{1}{2}$ and spin-${1}$ dimers with isotropic exchange coupling to the conduction electrons, recent experimental work emphasizes the importance of higher spins, single-ion anisotropy, anisotropic exchange and RKKY coupling as well as DM interactions \cite{Ruby2018, Kezilebieke2018, Choi2018, Beck2020, Ding2021, Kamlapure2019, Kuester2021,Liebhaber2021}. 

Here, we present a simple yet flexible approach to discuss dimers of quantum spins on superconductors. Sidestepping the substantial and rapidly forbidding numerical effort of full-scale NRG calculations, our approach focuses on the subgap excitations by limiting the substrate superconductor to a single site per adatom and conduction electron channel
(zero-bandwidth model \cite{Allub1981,Kirsanskas2015,Grove2018,Saldana2020,Oppen2021}). While this approach neglects Kondo renormalizations and effects associated with the spatial wave-function pattern of the YSR excitations, it is remarkably successful \cite{Oppen2021} in qualitatively reproducing the phase diagrams and excitation spectra of individual higher-spin adatoms, which were previously obtained by NRG \cite{Zitko2011}. We find that this remains true for adatom dimers. As detailed below, the approach qualitatively reproduces the phase diagrams and excitation spectra of spin-$\frac{1}{2}$ and spin-${1}$ dimers as obtained from NRG calculation in Ref.\ \cite{Zitko2011} and \cite{Yao2014}. This encourages us to apply the approach to models of the transition-metal and rare-earth systems used in the experiments on YSR dimers \cite{Ruby2018, Choi2018, Kezilebieke2018, Beck2020, Ding2021, Kamlapure2019, Kuester2021,Liebhaber2021}.

\begin{figure*}[t!]
         \centering
         \includegraphics[width=\textwidth]{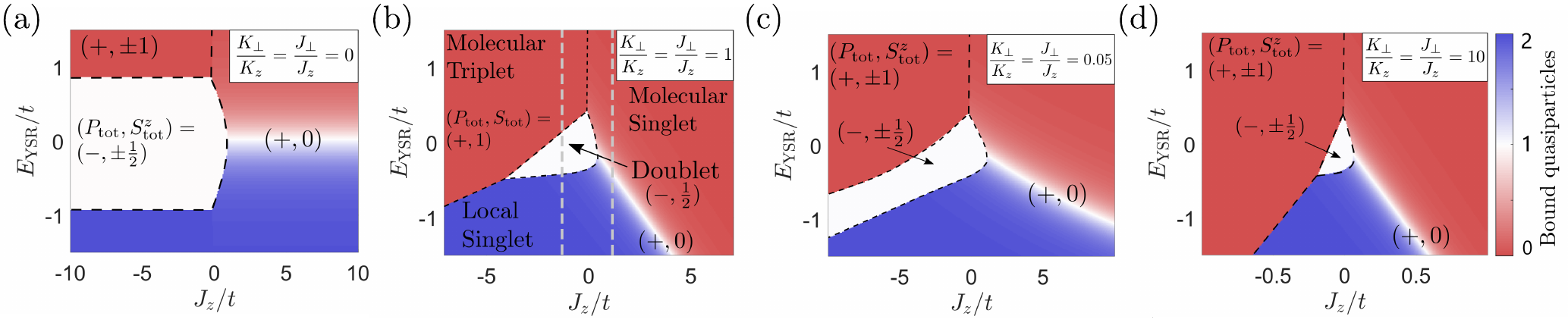}
	\caption{
	Phase diagrams of a spin-$\frac{1}{2}$ YSR dimer as a function of RKKY coupling $J_z$ and YSR energy $E_\text{YSR}$ for different anisotropies of the exchange coupling $K_\perp/K_z$. The anisotropy of the RKKY interaction is chosen to match the exchange anistropy, $K_\perp/K_z=J_\perp/J_z$. Phase diagrams are obtained from the zero-bandwidth Hamiltonian in Eq.\ (\ref{eq:model12}). The color scale (see scale bar) indicates the expectation value of the number of bound quasiparticles  $F$  as defined in Eq.\ (\ref{eq:nsqp}). Black dashed lines indicate phase boundaries, at which the spin and/or fermion parity quantum numbers of the ground state change discontinuously.
	(a)  Ising exchange with nonzero $K_z$ and $K_\perp=0$, corresponding to a classical-spin model of the adatom. Phase boundaries and crossovers essentially depend only on the sign of the RKKY coupling, reflecting the absence of screening of the adatom spin in the classical model. (b) Heisenberg exchange $K = K_\perp = K_z$. Phase boundaries and crossovers depend on the magnitude of the RKKY coupling. This phase diagram qualitatively reproduces the results of NRG simulations in Ref.\ \cite{Yao2014}. (c)  Dominant longitudinal and (d) dominant transverse anisotropic couplings as indicated in the panel. The singly-screened phase (white) reduces in extent as $K_\perp/K_z$ increases. Parameters: $(K_z,K_\perp)=100t( \cos\theta,\sin\theta)$, (a) $\theta=0$, (b) $\theta=\frac{\pi}{4}$, (c) $\theta=0.05$,  (d) $\theta=\frac{\pi}{2}-0.1$, $V=0.9(K_z+2K_\perp)/4$.}
	\label{fig_QuCl}
\end{figure*}  

There are important qualitative differences between the physics of classical and quantum spins on superconducting substrates. First, Kondo-like screening of the adatom spin is limited to quantum spins. Both classical and quantum adatom spins induce a quantum phase transition as their exchange coupling $K$ to the conduction electrons increases \cite{Sakurai1970}. At weak coupling, the ground state of the superconductor is fully paired (even fermion parity). Beyond a critical coupling, the adatom binds a quasiparticle (odd fermion parity). However, only for quantum spins, this binding of a quasiparticle is associated with a change in the ground-state multiplicity and thus with Kondo-like screening of the adatom spin. In dimers, this screening abruptly alters the RKKY energy at the quantum phase transition \cite{Steiner2021,Liebhaber2021}, leading to rich physics of quantum YSR dimers. Second, half-integer and integer quantum spins can behave in qualitatively different ways due to the presence or absence of Kramers degeneracies. This leads to characteristic differences in their Kondo effects \cite{Otte2008} and we find related distinctions for quantum YSR dimers. 

This paper is organized as follows. Section \ref{sec:12} discusses YSR spin-$\frac{1}{2}$ dimers, contrasting classical and quantum spins. Motivated by transition-metal and rare-earth systems, Sec.\ \ref{sec_higher_spins} extends the discussion to dimers of higher-spin adatoms, accounting for  single-ion anisotropy and Dzyaloshinsky-Moriya interactions. We conclude in Sec.\ \ref{sec:con}. In an effort to focus the main text on the principal physical arguments, we delegate some technical details as well as some additional considerations to appendices. 

\section{Spin-$\frac{1}{2}$ dimers}  
\label{sec:12}

\subsection{Monomers and screening}

The qualitatively different screening behavior of classical and quantum adatom spins can be understood by considering a spin-$\frac{1}{2}$ monomer within the zero-bandwidth model 
\begin{equation}
   H = \Delta (c^\dagger_{\uparrow}      
       c^\dagger_{\downarrow}  +   \mathrm{h.c.} ) + c_{\sigma}^\dagger  [V \delta_{\sigma\sigma'} +   \mathbf{S}
       \cdot \hat{K} \cdot \mathbf{s}^{\phantom{\dagger}}_{\sigma\sigma'}] c^{\phantom{\dagger}}_{\sigma'}
\label{eq:monomodel}
\end{equation}
(see App.\ \ref{app:spin12_monomer} and \cite{Oppen2021} for further discussion). Here, $\Delta$ is the pairing amplitude of the superconducting site coupled to the adatom spin $\mathbf{S}$ through potential scattering $V$ and antiferromagnetic exchange interaction $\hat{K}=\textrm{diag}(K_\perp,K_\perp,K_{z})$. On the superconducting site, conduction electrons of spin $\sigma$ are annihilated by $c_{\sigma}$, and $\mathbf{s} =\frac{1}{2}\boldsymbol{\sigma}$ in terms of the vector of Pauli matrices $\boldsymbol{\sigma}$. Summation over repeated spin indices is implied. 

Within models of classical spins, one assumes that the adatom spin $\mathbf{S}$ is aligned along, say, the $z$ direction, so that it couples only to a single component of the conduction-electron spin (density) $c^\dagger_\sigma \mathbf{s}^{\phantom{\dagger}}_{\sigma\sigma'} c^{\phantom{\dagger}}_{\sigma'}$ and there are no transverse spin couplings. Within the model of Eq.\ (\ref{eq:monomodel}), this corresponds to Ising exchange coupling $\hat{K}=\mathrm{diag}(0,0,K_{z})$. In contrast, the quantum nature of the adatom spin plays a role as soon as the transverse spin couplings are nonzero, $K_\perp \neq 0$, as is the case, for instance, for Heisenberg coupling $\hat{K}=\mathrm{diag}(K,K,K)$. 

Regardless of $K_\perp$, the monomer ground state exhibits a quantum phase transition with increasing exchange coupling. It is fully paired with a free adatom spin at weak exchange coupling and binds a quasiparticle at strong coupling. The weak-coupling state $\ket{\Uparrow/\Downarrow,\text{BCS}}$ is a direct product of a free impurity spin ($\ket{\Uparrow/\Downarrow}$) and a paired electronic ground state ($\ket{\text{BCS}}$) with even fermion parity, and takes the same form for classical and quantum spins. In contrast, the strong-coupling states of classical and quantum spins differ in their screening properties. In the classical case, the monomer continues to have two degenerate ground states, namely the odd-fermion-parity states $\ket{\Uparrow,\downarrow}$ and $\ket{\Downarrow,\uparrow}$. Consequently, the quantum phase transition leaves the impurity-spin state unaffected and thus unscreened.  
For quantum spins, the nonzero transverse exchange coupling $K_\perp$ lifts the degeneracy between $\ket{\Uparrow,\downarrow}$ and $\ket{\Downarrow,\uparrow}$ and the singlet state $\ket{s} = \ket{\Uparrow,\downarrow}-\ket{\Downarrow,\uparrow}$ becomes the unique strong-coupling ground state. Now, the impurity spin no longer points along a preferred direction and is thus screened by the conduction electrons (see also App.\ \ref{app:spin12_monomer}). 

The different phases of the monomer can in general be labeled by the fermion parity $P=(-1)^{\sum_\sigma c_\sigma^\dagger c_\sigma^{\phantom{\dagger}}}$ as well as the magnitude and/or projection of the effective spin
\begin{equation}
    \mathbf{S}_\mathrm{eff} = \mathbf{S} + c_\sigma^\dagger \mathbf{s}^{\phantom{\dagger}}_{\sigma\sigma'}c_{\sigma'}^{\phantom{\dagger}},
    \label{eq_seff}
\end{equation}
depending on the degree of spin rotation symmetry \cite{Oppen2021}. The excitation energy 
\begin{equation}
    E_\mathrm{YSR} = E_o-E_e.
\end{equation}
of the YSR state is the energy difference of the lowest monomer states in the odd and even-fermion-parity sectors ($E_o$ and $E_e$, respectively). With this definition, $E_\mathrm{YSR}$ is positive in the weak-coupling phase and negative in the strong-coupling phase, and takes on the value
\begin{equation}
   E_\mathrm{YSR} = \sqrt{\Delta^2+V^2}-\frac{1}{4}(K_z+2K_\perp) 
\end{equation}
for a spin-$\frac{1}{2}$ monomer. 

The zero-bandwidth approximation fails to account for the quasiparticle continuum and can thus only be expected to describe deep subgap states. We account for this limitation by assuming large $\Delta$, $K$, and $V$ in such a way that the YSR energy $E_\mathrm{YSR}$  and the dimer couplings remain small by comparison. In particular, this assures that $E_\mathrm{YSR}$ is well within the gap. This assumption will be made throughout this paper, in both the numerical and the analytical calculations.  

\subsection{Dimer phase diagrams}
\label{sec_dimer_12}

The distinctly different screening properties of classical and quantum spins have important ramifications for the phase diagram of dimers. This can already be illustrated for a spin-$\frac{1}{2}$ dimer within the zero-bandwidth model   
\begin{eqnarray}
   &&H = \sum_{j=1}^2 \Delta (c^\dagger_{j\uparrow}      
       c^\dagger_{j\downarrow}  +   \mathrm{h.c.} ) -t  [c^{\dagger}_{1\sigma}
       c^{\phantom{\dagger}}_{2\sigma} + \mathrm{h.c.}]   \cr
   && \,\,\,\,\,\,\,
   +  \sum_{j=1}^2 c_{j\sigma}^\dagger  [V \delta_{\sigma\sigma'} +   \mathbf{S}_{j}
       \cdot \hat{K} \cdot \mathbf{s}^{\phantom{\dagger}}_{\sigma\sigma'}] c^{\phantom{\dagger}}_{j\sigma'}  + \mathbf{S}_1 \cdot \hat{J} \cdot  \mathbf{S}_{2}. \,\,\,\,
\label{eq:model12}
\end{eqnarray}
Here, the adatom spins $\mathbf{S}_j$ ($j=1,2$) are coupled to separate superconducting sites ($c_{j\sigma}$). Hybridization of the YSR states is incorporated through intersite hopping of strength $t$. The RKKY interaction $\hat{J} = \textrm{diag}(J_\perp,J_\perp,J_{z})$ is incorporated explicitly as it is mediated by the quasiparticle continuum, which is not accounted for within the zero-bandwidth model. Due to the oscillatory dependence of the RKKY interaction, strength and sign of $\hat{J}$ depend on the distance between the adatoms. 

\begin{figure}[t!]
       \centering
     \includegraphics[width=0.4\textwidth]{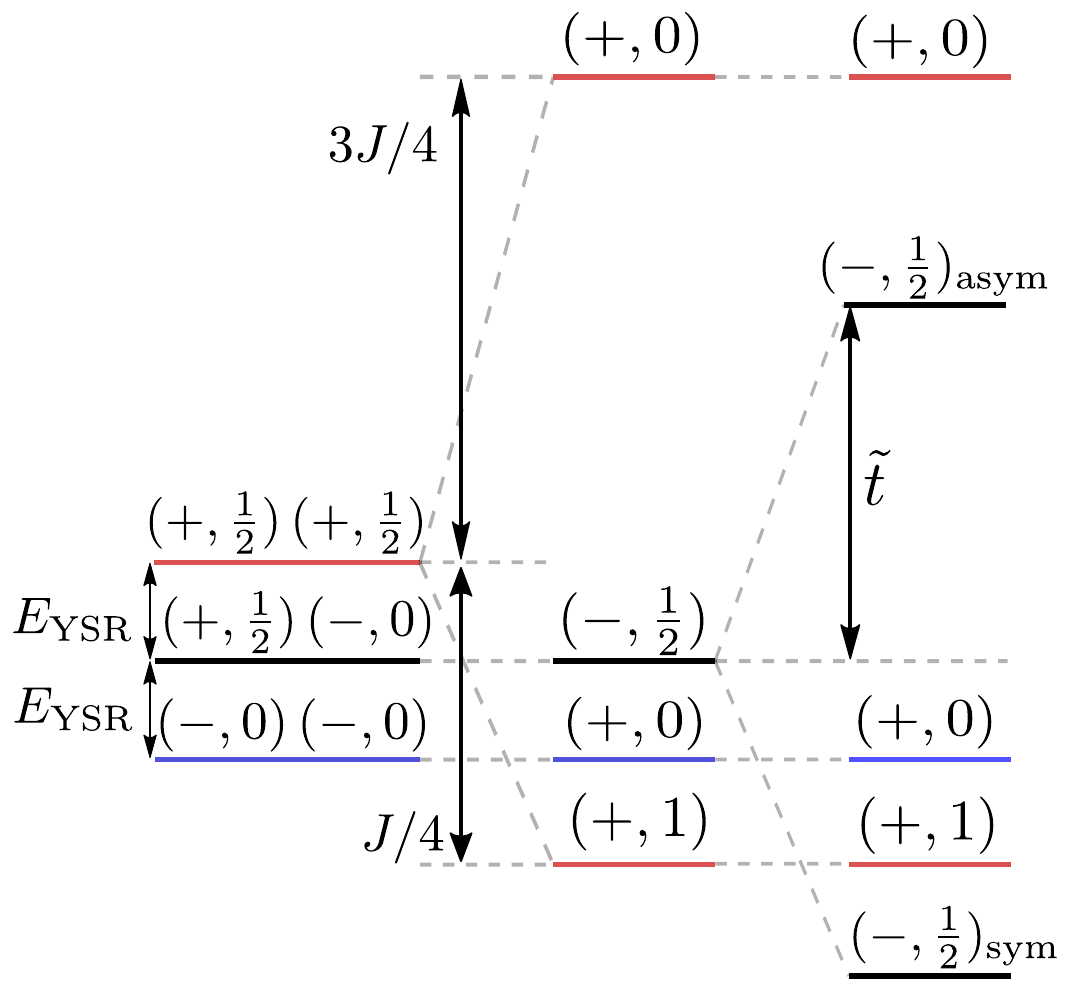}
     \caption{Illustrative level scheme of spin-$\frac{1}{2}$ dimers with isotropic exchange and ferromagnetic RKKY coupling $(J<0)$. The low-energy spectrum of the uncoupled dimer (left) is labeled by the fermion parities $P_j$ and effective spins $S_{\mathrm{eff},j}$ of the monomers as $(P_1,S_{\mathrm{eff},1})(P_2,S_{\mathrm{eff},2})$. For $E_\text{YSR}<0$ the local singlet state ($F \simeq 2$, blue) is the ground state. Nonzero RKKY interaction (center) couples the monomer states into states of total parity and spin $(P_\text{tot},S_\text{tot})$. This affects only the unscreened state ($F\simeq 0$, red), which splits into molecular singlet and triplet. For sufficiently large $|J|$, the molecular triplet becomes the ground state. Finally, hybridization of the YSR states splits the odd-fermion-parity states ($F= 1$, black) into symmetric and antisymmetric states. For sufficiently large hybridization $\tilde{t}$ this leads to the singly-screened ground state.}
	\label{fig_doublet12}
\end{figure}  

To characterize the phases of the model in Eq.\ (\ref{eq:model12}), we exploit the symmetries of the system.
The superconducting pairing breaks particle-number conservation, but conserves the overall fermion parity 
\begin{equation}\label{eq:p_tot}
   P_\mathrm{tot} = (-1)^{\sum_\sigma (c^{\dagger}_{1\sigma}c^{\phantom{\dagger}}_{1\sigma}+c^{\dagger}_{2\sigma}c^{\phantom{\dagger}}_{2\sigma})}. 
\end{equation}
Provided that the model retains spin rotation symmetry about the $z$-axis, the projection ${S}^z_{\mathrm{tot}}$ of the total spin 
\begin{equation}\label{eq:S_tot}
 \mathbf{S}_\mathrm{tot}=\mathbf{S}_\mathrm{1}+\mathbf{S}_\mathrm{2}+ \sum_{j}c_{j\sigma}^\dagger  \mathbf{s}^{\phantom{\dagger}}_{\sigma\sigma'} c^{\phantom{\dagger}}_{j\sigma'} 
\end{equation}
is also a conserved quantity. For the special case of Heisenberg exchange and RKKY interactions, the model has full spin rotation symmetry and we can further classify phases according to ${S}_\mathrm{tot}$. Finally, we can label the dimer phases by their spatial parity $\Sigma$ (with $\Sigma^2 = 1$), which interchanges the monomers as defined by 
\begin{equation}
    \Sigma c^{\dagger}_{1,\sigma}  \Sigma = c^{\dagger}_{2,\sigma} \quad ,\quad  \Sigma \mathbf{S}_1 \Sigma = \mathbf{S}_2. 
    \label{eq:SigmaDef}
\end{equation}
Note that this operation also exchanges the fermions, which gives rise to an additional minus sign when both adatom spins are screened.

\begin{figure*}[t!]
         \centering
 \begin{minipage}[b]{0.45\textwidth}
         \centering
     \includegraphics[width=\textwidth]{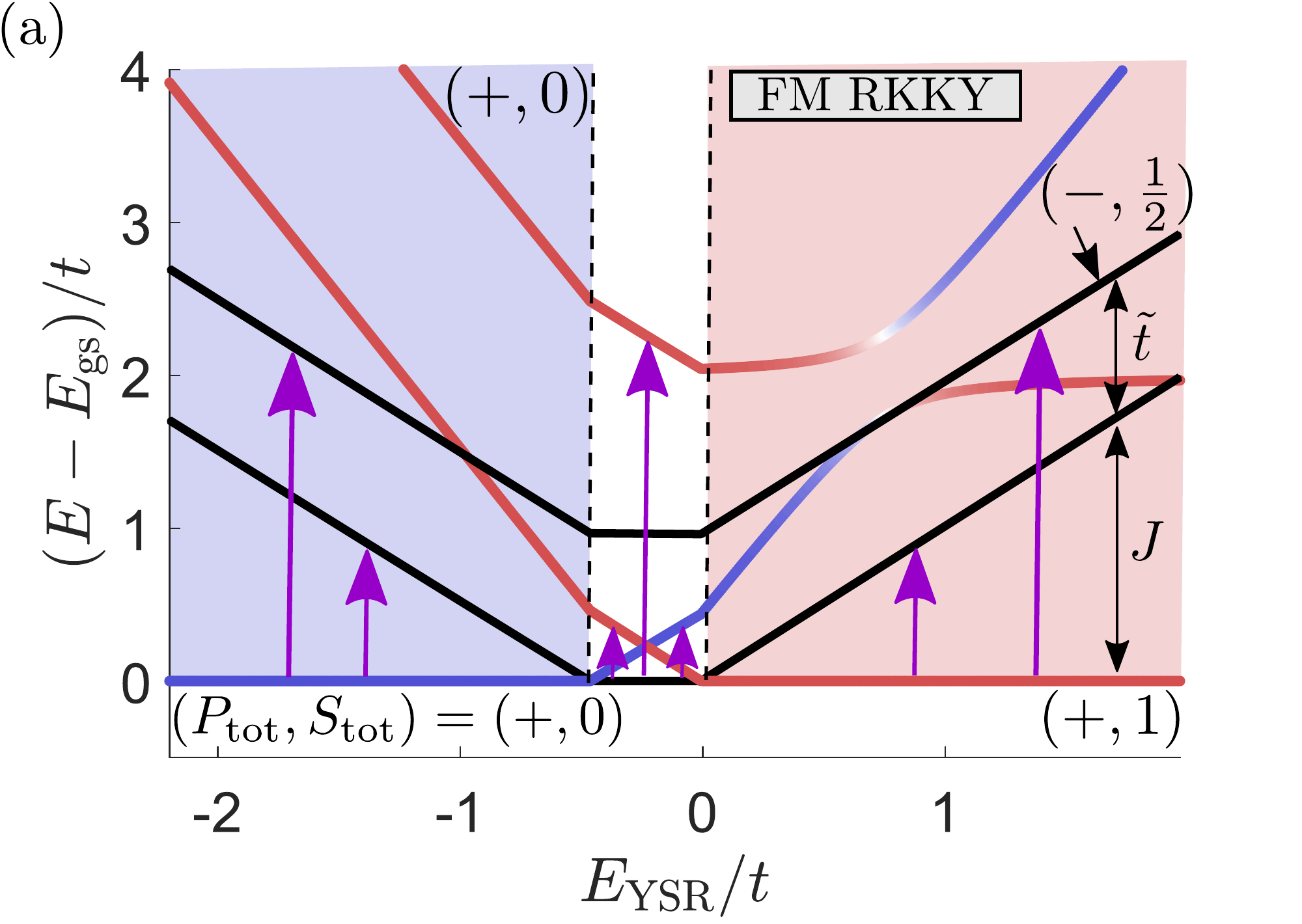}
          \end{minipage}
     \hfill
    \begin{minipage}[b]{0.49\textwidth}
         \centering
         \includegraphics[width=\textwidth]{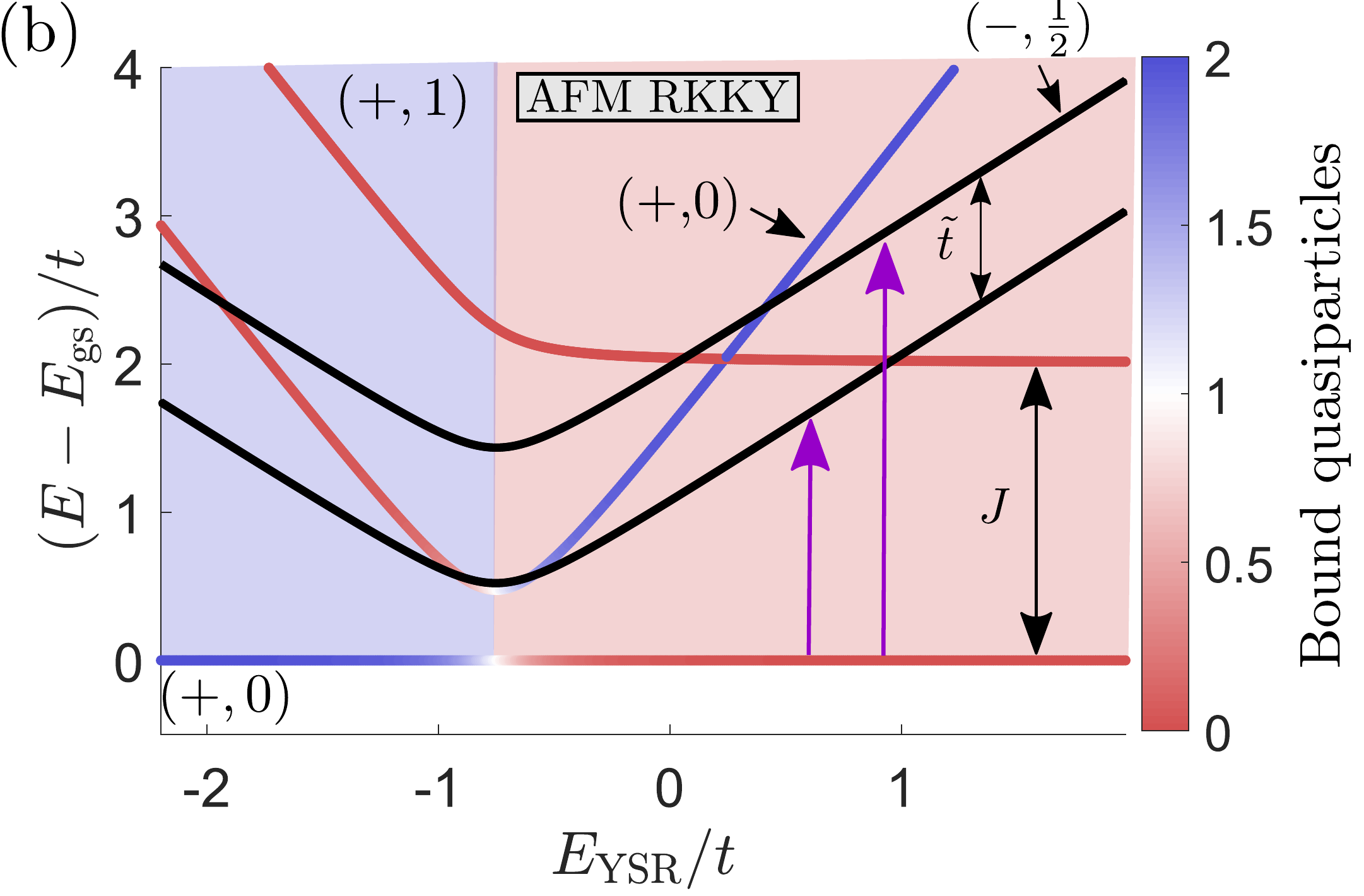}
     \end{minipage}
	\caption{Excitation spectra of a spin-$\frac{1}{2}$ dimer as a function of YSR energy $E_\text{YSR}$ for isotropic (a) ferromagnetic (FM) and (b) antiferromagnetic (AFM) RKKY coupling $J$. Tunneling excitations (purple arrows) flip fermion parity and change total spin by $\pm \frac{1}{2}$. In the even-fermion-parity phases, tunneling is possible into the symmetric and antisymmetric doublets. In the doublet phase, all (low-energy) even parity states are in principle accessible by tunneling, resulting in three peaks (or more for anisotropic RKKY interaction, see Fig.\ \ref{fig_spectral_fun_spin_12}). Parameters as in Fig.\ \ref{fig_QuCl}(b) with fixed RKKY coupling $J=-2t$ (a) and $J=2t$ (b), as indicated by gray dashed lines in Fig.\ \ref{fig_QuCl}(b) . }
	\label{fig_excitation12}
\end{figure*}  

The quantum numbers $P_\mathrm{tot}$, ${S}^z_{\mathrm{tot}}$, and $\Sigma$ can be used to classify the model's phases. We also find it useful to consider the expectation value of 
\begin{align}
    F=\sum\limits_{j=1,2}\left(
    c^{\dagger}_{j\uparrow}c^{\phantom{\dagger}}_{j\uparrow}-c^{\dagger}_{j\downarrow}c^{\phantom{\dagger}}_{j\downarrow}\right)^2,
    \label{eq:nsqp}
\end{align}
which is a proxy for the number of bound quasiparticles. It should be noted, however, that for nonzero YSR hybridization $t$, this is not a conserved quantum number due to the presence of pairing correlations in the model. While ground states with different quantum numbers define phases of the quantum YSR dimer, we refer to the unscreened ($F\simeq 0$) or doubly-screened ($F\simeq 2$) parts of the phase diagram as regions. 

Figure \ref{fig_QuCl}(a) and (b) show dimer phase diagrams as a function of the RKKY interaction $J_z$ and the YSR energy $E_\mathrm{YSR}$ for Ising coupling (classical spins) and Heisenberg coupling (quantum spins), respectively. The most striking difference is that phase boundaries and crossovers in the classical phase diagram [Fig.\ \ref{fig_QuCl}(a)] essentially depend only on the sign, but not on the magnitude of the RKKY interaction $J$. In contrast, the magnitude of the RKKY coupling is an important parameter in the quantum phase diagram [Fig.\ \ref{fig_QuCl}(b)].

This difference arises as follows. The phase boundaries correspond to lines in the phase diagram, along which states with different total spin are degenerate. For a classical impurity spin, the quantum phase transition does not affect the impurity-spin state. Consequently, it leaves the RKKY energy of the dimer unchanged, which will then cancel from the energy balance governing phase boundaries and crossovers. In contrast, for quantum spins, the impurity spin is fully screened in the strong-coupling state. Thus, only the unscreened phases benefit from the RKKY interaction, while the RKKY interaction energy vanishes for the phases in which one or both spins are screened. Now, the RKKY interaction enters into the energy balance governing phase boundaries and crossovers. 

The phase diagram of a quantum spin-$\frac{1}{2}$ dimer with Heisenberg interactions was also computed in Ref.\ \cite{Yao2014}, using the numerical renormalization group (NRG) including the full quasiparticle continuum of the substrate superconductor. Remarkably, the phase diagram of the zero-bandwidth model of Eq.\ (\ref{eq:model12}) in Fig.\ \ref{fig_QuCl}(b) qualitatively reproduces the NRG phase diagram. 
We now discuss the phase diagram in Fig.\ \ref{fig_QuCl}(b) for isotropic (Heisenberg) exchange and RKKY coupling in more detail (see also App.\ \ref{app:spin12_dimer_iso}). For sufficiently large $E_\mathrm{YSR}$, both adatom spins are unscreened ($F \simeq 0$). For ferromagnetic RKKY coupling ($J<0$), the ground state is a molecular triplet, e.g., $\ket{\Uparrow,\textrm{BCS}}_1 \otimes \ket{\Uparrow,\textrm{BCS}}_2$, with quantum numbers $(P_\mathrm{tot},S_\mathrm{tot})=(+,1)$. For antiferromagnetic RKKY interactions ($J>0$), the unscreened impurity spins couple into a molecular singlet $\ket{\Uparrow,\textrm{BCS}}_1 \otimes \ket{\Downarrow,\textrm{BCS}}_2-\ket{\Downarrow,\textrm{BCS}}_1 \otimes \ket{\Uparrow,\textrm{BCS}}_2$, so that $(P_\mathrm{tot},S_\mathrm{tot})=(+,0)$. For large and negative $E_\mathrm{YSR}$, the adatom spins are individually screened ($F \simeq 2$) and the dimer has a local-singlet ground state, $\ket{s}_1 \otimes \ket{s}_2$. This state also has $(P_\mathrm{tot},S_\mathrm{tot})=(+,0)$, so that the molecular singlet evolves continuously into the local singlet phase as $E_\mathrm{YSR}$ is reduced. The absence of a sharp phase transition between these ground states reflects that the pairing term in the model in Eq.\ (\ref{eq:model12}) breaks particle-number conservation. At large and negative RKKY coupling, there is a direct transition between the molecular triplet and the local-singlet phases, with a corresponding change in $S_\mathrm{tot}$.  

For weak RKKY coupling and small $\abs{E_\mathrm{YSR}}$, there is an odd-fermion-parity phase with half-integer total spin,  $(P_\mathrm{tot},S_\mathrm{tot}) =(-,\frac{1}{2})$. The doublet ground state of this phase, $\ket{s}_1 \otimes \ket{\Uparrow/\Downarrow,\textrm{BCS}}_2 + \ket{\Uparrow/\Downarrow,\textrm{BCS}}_1 \otimes \ket{s}_2$, emerges when the hybridization splitting of the dimer states with one screened adatom is large enough to offset the cost in YSR energy.  

Figure \ref{fig_doublet12} shows level diagrams for the RKKY and hybridization splittings, illustrating the mechanisms governing the phase diagram in Fig.\ \ref{fig_QuCl}(b). While the ferromagnetic RKKY coupling favors the molecular triplet, a sufficiently large YSR hybridization $t$ can lower the energy of the doublet with half-integer spin to become the ground state. Note that we use $2\tilde t$ to denote the actual energy splitting of the singly-screened states due to the hybridization $t$. 

The phase diagrams in Fig.\ \ref{fig_QuCl}(c) and (d) 
for, respectively, predominantly longitudinal and transverse exchange and RKKY couplings deviate qualitatively from the isotropic Heisenberg case. Here, we take both the exchange coupling $\hat{K}$ and the RKKY interaction $\hat{J}$ to have the same anisotropy (i.e., $K_\perp/K_z = J_\perp/J_z$). For dominant longitudinal coupling, $K_{\perp} \ll K_z$, Fig.\ \ref{fig_QuCl}(c), the doublet phase continues to form a stripe as in the Ising case, albeit with boundaries that depend on the RKKY coupling. Beyond a critical value of $K_{\perp}$, the doublet phase forms an island as in the Heisenberg case. 
We note that classical behavior with phase boundaries approximately independent of $J$ is recovered only for $K_\perp\ll J,t$. For dominant transverse couplings, $K_{\perp} \gg K_z$, Fig.\ \ref{fig_QuCl}(d), the doublet phase remains limited to small RKKY couplings as in the isotropic Heisenberg case.  

\subsection{Dimer excitation spectra}

Figure \ref{fig_excitation12} shows representative excitation spectra for Heisenberg couplings along the dashed lines in Fig.\ \ref{fig_QuCl}(b). These provide further confirmation that the dashed lines cross true quantum phase transitions on the ferromagnetic side, as indicated by the level crossings, while there are only anticrossings indicative of crossovers on the antiferromagnetic side. The excitation spectra are in excellent qualitative agreement with corresponding results of NRG calculations in Ref.\ \cite{Yao2014}, further supporting the usefulness of the zero-bandwidth model \footnote{\citet{Yao2014} assumed a particle-hole symmetric substrate which gives rise to a level crossing between the local-singlet and molecular-singlet states. As we do not assume this additional symmetry, we observe the more generic avoided crossing.}.

\begin{figure}[t!]
         \centering
     \includegraphics[width=0.45\textwidth]{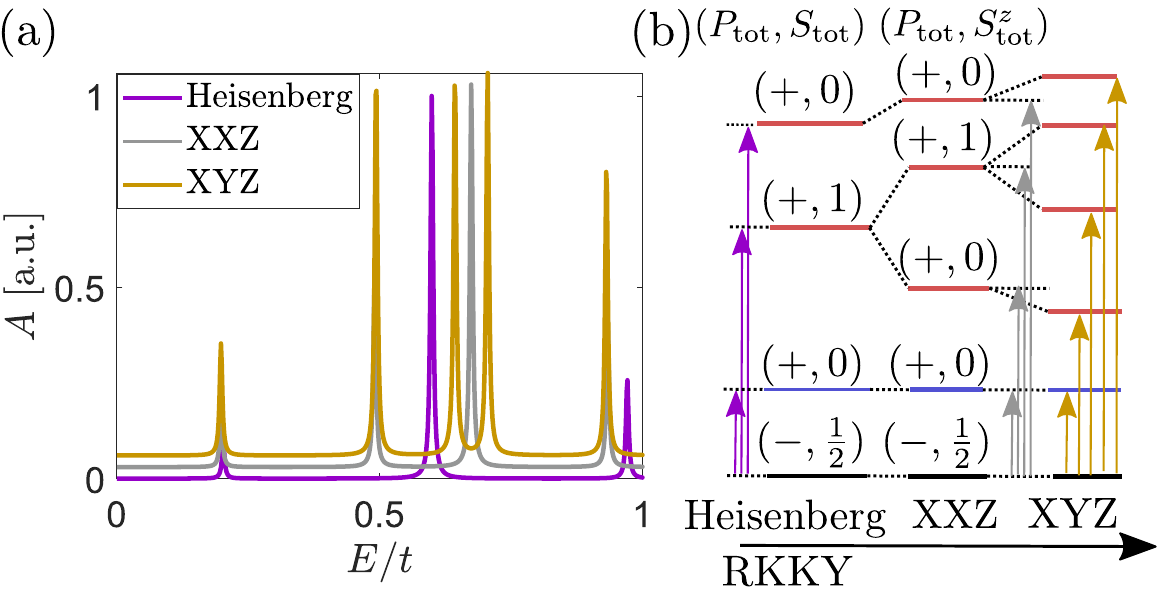}
	\caption{Tunneling spectroscopy of spin-$\frac{1}{2}$ dimers in the doublet phase with isotropic exchange and ferromagnetic RKKY coupling. (a) Spectral functions for different symmetries of the RKKY coupling (see legend; offset for clarity). (b) Level schemes (not to scale) for Heisenberg, XXZ and XYZ RKKY coupling, emphasizing the increase in the number of resonances as the spin rotation symmetry is reduced. Parameters: $E_\mathrm{YSR}=-0.2t$, $\Delta = 10^6 t$, $V=
	3\Delta$, $J=-0.3t$, Heisenberg: $\hat{J}=J\,\mathrm{diag}(1,1,1)$, XXZ: $\hat{J}=J\sqrt{3/2}\,\mathrm{diag}(1,1,0)$, XYZ: $\hat{J}=J\,\mathrm{diag}(\sqrt{2},1,0)$.}
	\label{fig_spectral_fun_spin_12}
\end{figure}  

To connect to tunneling experiments, we also consider the spectral function 
\begin{align}
    A(E)=\sum\limits_{j=1}^2\sum\limits_ {\sigma \lambda} \big[&|\langle\lambda|c^\dagger_{j,\sigma}|\mathrm{gs}\rangle|^2\,\delta\left(E-E_\lambda+E_\mathrm{gs}\right) +
    \notag
    \\
    &|\langle\lambda|c_{j,\sigma}|\mathrm{gs}\rangle|^2\,\delta \left(E+E_\lambda-E_\mathrm{gs}\right) \big],
\end{align}
where $|\mathrm{gs}\rangle$ denotes the ground state and $|\mathrm{\lambda}\rangle$ the excited states with opposite fermion parity. Spin rotation symmetries give rise to selection rules for the matrix elements. In Fig.\ \ref{fig_spectral_fun_spin_12}, we show $A(E)$ for the doublet phase. For full spin rotation symmetry, there are three peaks, corresponding to transitions into the  molecular singlet and triplet as well as the local singlet. Considering RKKY couplings with XXZ and XYZ symmetry, the triplet peak splits into two and then three resonances. Thus, depending on the degree of symmetry, there are between three and five peaks in the doublet phase. For even-parity ground states, there are only two peaks in $A(E)$, regardless of the degree of spin rotation symmetry. These correspond to the symmetric and antisymmetric doublets, which do not split further due to Kramers degeneracy. Thus, quantum phase transitions generically change the number of resonances observed in tunneling experiments.

\begin{figure*}[t]
 \centering  
 \includegraphics[width=\textwidth]{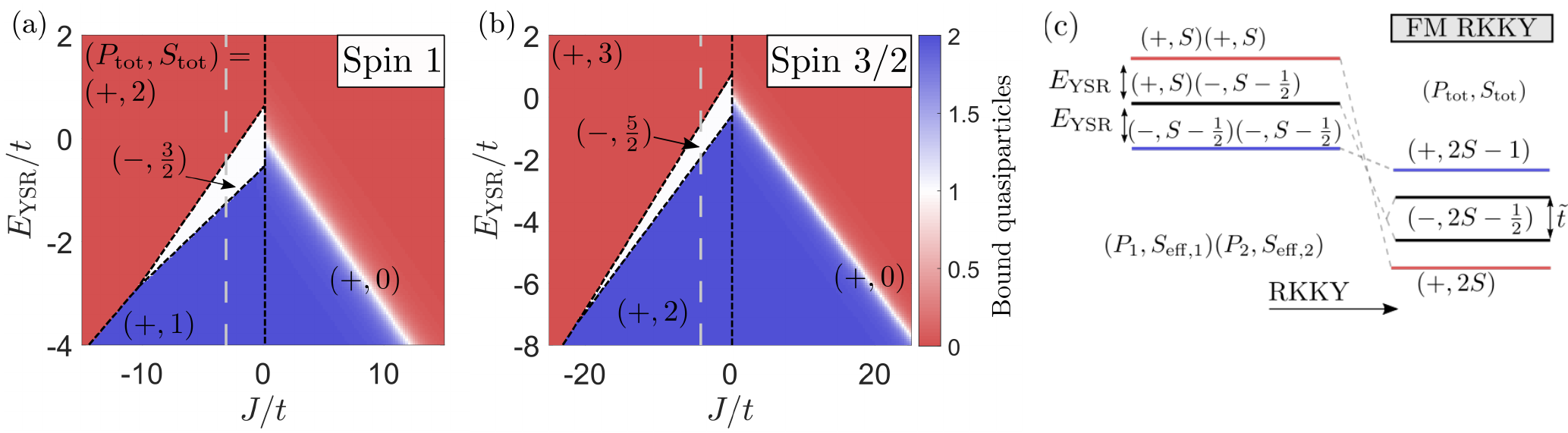}
     \caption{Effect of Heisenberg RKKY coupling on YSR dimers with higher spins $S$. (a),(b) Phase diagrams as a function of isotropic RKKY coupling $J$ and YSR energy $E_\mathrm{YSR}$ for $S$ as indicated in the panel.  As for spin-$\frac{1}{2}$ dimers, RKKY coupling stabilizes unscreened dimers with $F \simeq 0$ (red).  
     Unlike spin-$\frac{1}{2}$ dimers, the phase boundary between the approximately doubly-screened $S_\mathrm{tot}=2S-1$ (blue, $F \simeq 2$) and  singly-screened  $S_\mathrm{tot}=2S-\frac{1}{2}$ phases  (white, $F  = 1$) is no longer approximately constant as partially screened dimers gain RKKY energy, see (c). As $S$ increases, the phase boundaries on the ferromagnetic side become more and more parallel. Gray dashed lines indicate value of the RKKY coupling in Fig.\ \ref{fig_excitation_higherspins}. 
     (c) Level scheme for higher spins. RKKY coupling shifts the energies not only of the unscreened dimer, but also of the singly and doubly screened dimer. Parameters: $V=2\Delta$, $\Delta = 10^6 t$.}
 \label{fig:higher_spin_phase_diagrams_iso}
\end{figure*}

\section{Higher spins} 
\label{sec_higher_spins}

\subsection{Isotropic RKKY coupling}
\label{subsec_higher_spins_RKKY}

YSR dimers based on transition-metal and rare-earth systems typically involve higher-spin adatoms. In many cases, higher spins effectively behave more classically. Remarkably, in the present problem, the phase diagrams for classical and quantum spins remain qualitatively distinct even as $S\to\infty$. This is a direct consequence of the difference in the screening properties. Although a single conduction-electron channel screens higher quantum spins merely from $S$ to $S-\frac{1}{2}$, we find that the change in RKKY energy associated with a quantum phase transition to the screened state approaches a nonzero constant and thus remains relevant even for large $S$.

Figure \ref{fig:higher_spin_phase_diagrams_iso} shows representative phase diagrams for larger adatom spins as a function of the RKKY coupling $J$ and the YSR energy $E_\mathrm{YSR}$. Both the exchange coupling $K$ and the RKKY coupling $J$ are assumed isotropic, so  that for general impurity spin $S$, the YSR energy of the monomer is given by (see App.\ \ref{app:higher_spin_dimer_iso_mono})
\begin{equation}
  E_\mathrm{YSR}  = \sqrt{\Delta^2 + V^2} - \frac{S+1}{2}K.
  \label{eq:EYSR_iso}
\end{equation}
As appropriate for isotropic couplings, we label  phases by their total spin $S_{\textrm{tot}}$. The phase diagrams for adatom spins $S = 1$ and $S = \frac{3}{2}$ in Fig.\ \ref{fig:higher_spin_phase_diagrams_iso} exhibit four phases, one more than for the spin-$\frac{1}{2}$ dimer. For a single conduction-electron channel per adatom, higher adatom spins can only be partially screened. Thus, even in the doubly-screened region, they will couple to different total spins for ferromagnetic and antiferromagnetic RKKY coupling. This introduces a phase boundary at $J=0$ and negative $E_\mathrm{YSR}$, which is absent for a dimer of fully screened $S=\frac{1}{2}$ adatoms. 

Apart from this additional phase boundary, the phase diagrams for higher spins are similar in appearance to the phase diagram for $S=\frac{1}{2}$. In particular, nonzero RKKY coupling of either sign favors the unscreened phases. At the same time, the half-integer spin phase [white region in Figs.\ \ref{fig:higher_spin_phase_diagrams_iso}(a) and (b)]
has not only a different ground-state multiplicity (doublet for $S=\frac{1}{2}$, quartet for $S=1$, and sextet for $S=\frac{3}{2}$), but also appears in a differently shaped parameter region. While the lower boundary is only weakly dependent on $J$ for $S=\frac{1}{2}$ [Fig.\ \ref{fig_QuCl}(b)], it exhibits a pronounced linear $J$ dependence for higher spins. For $S=\frac{1}{2}$, both the singly-screened and the doubly-screened regions have zero RKKY energy and $J$ does not enter the energy balance determining their phase boundary. In contrast, singly and doubly-screened regions have nonzero -- and different -- RKKY energies for higher spins [see Fig.\ \ref{fig:higher_spin_phase_diagrams_iso}(c)], leading to a $J$-dependent phase boundary. 

We note in passing that Fig.\ \ref{fig:higher_spin_phase_diagrams_iso} shows phase diagrams for $V\sim\Delta$. At small $\abs{J}$, the detailed structure of the phase diagram actually depends in some detail on the ratio $V/\Delta$, see App.\ \ref{app:higher_spin_dimer_iso_phase} for further discussion. In particular, there can be additional spin transitions which have no analog for classical adatom spins. 

\begin{figure*}[t!]
         \centering
   \begin{minipage}[b]{0.45\textwidth}
         \centering
     \includegraphics[width=\textwidth]{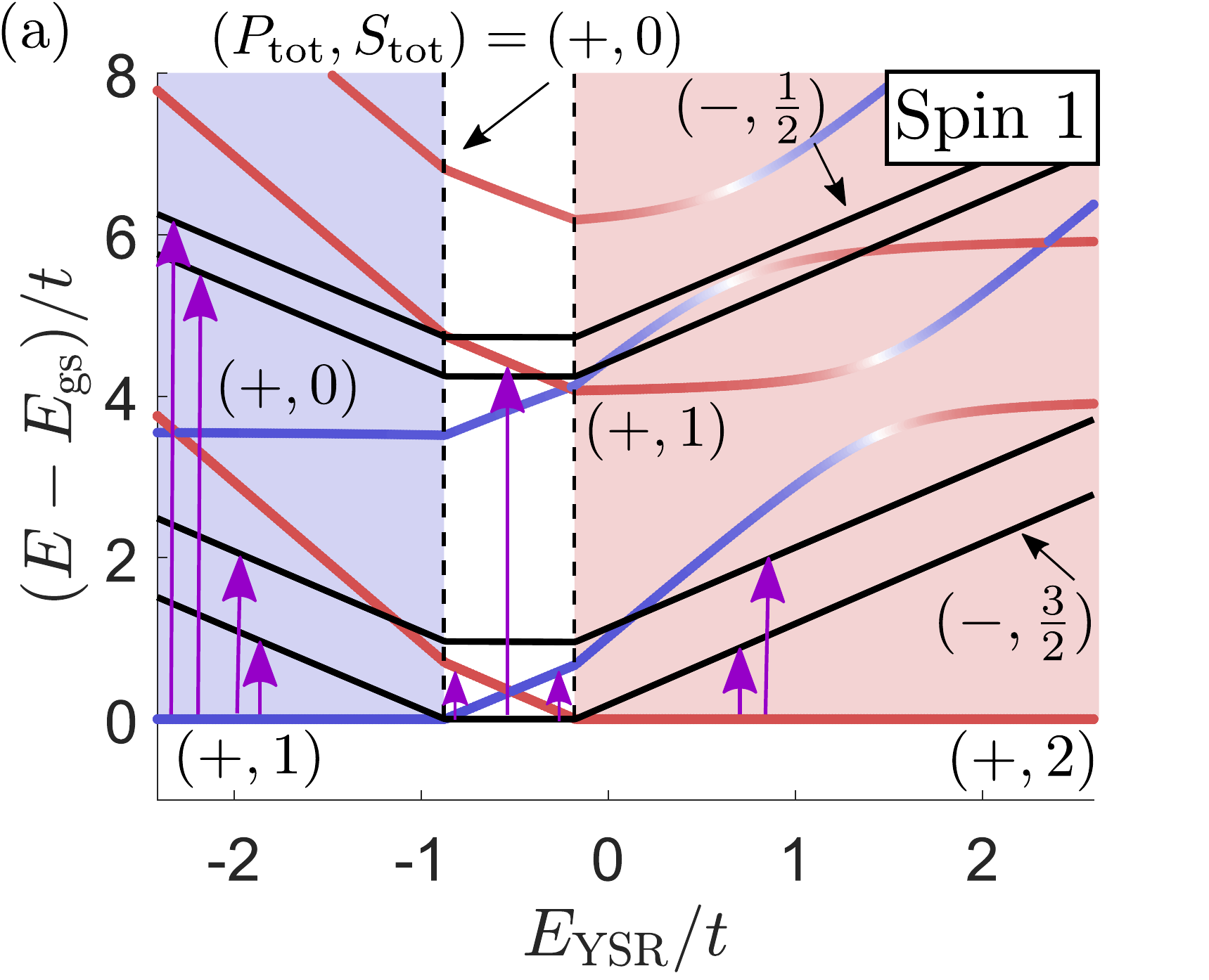}
     \end{minipage}
     \hfill
    \begin{minipage}[b]{0.46\textwidth}
         \centering
     \includegraphics[width=\textwidth]{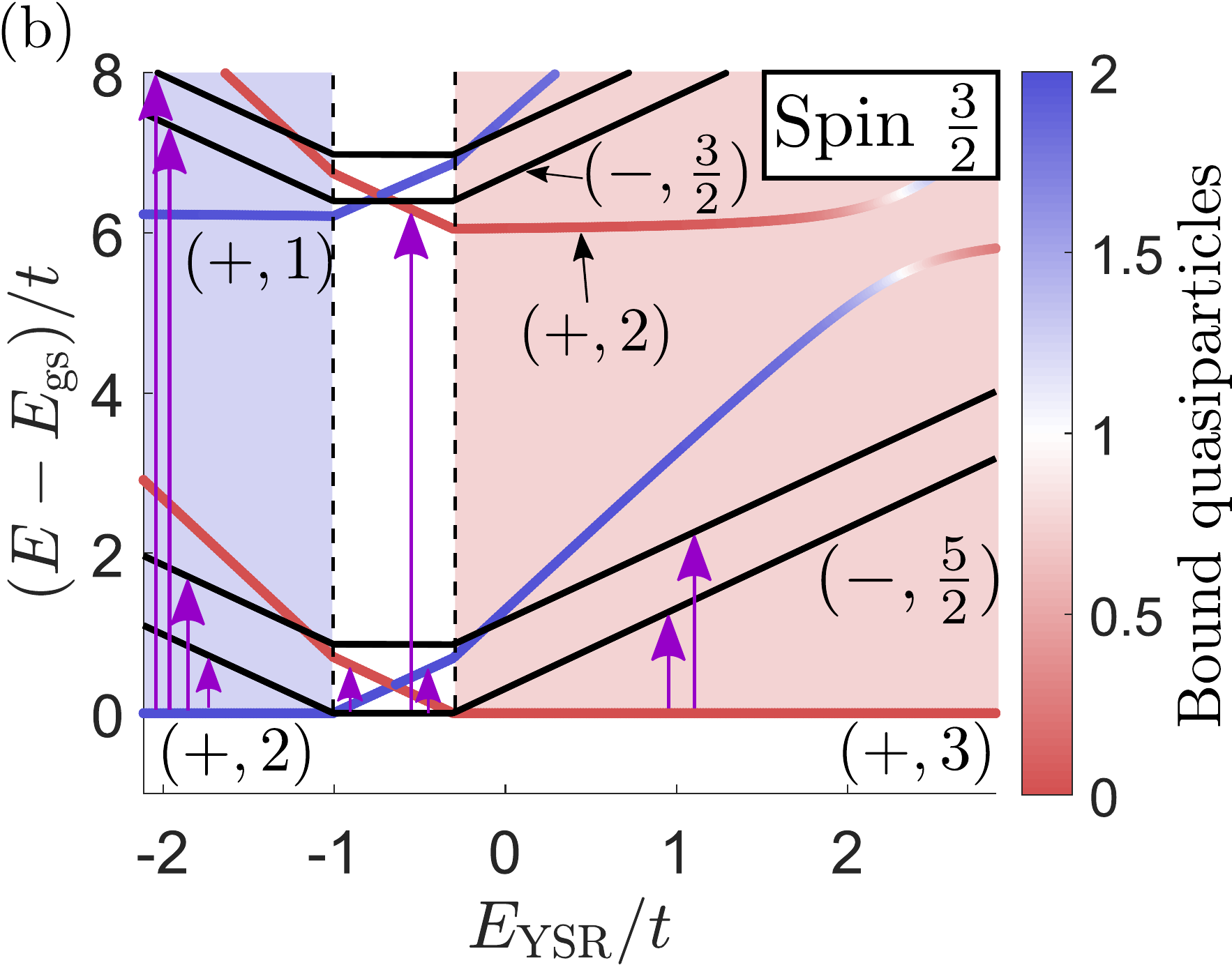}
     \end{minipage}
	\caption{Excitation spectra of YSR dimers for adatom spins (a) $S=1$ and (b) $S=\frac{3}{2}$ with isotropic ferromagnetic RKKY coupling. Allowed tunneling excitations (purple arrows) are determined by the selection rules $P_\mathrm{tot} P_\mathrm{tot}' = -1$ and $S^z_\mathrm{tot} - (S^z_\mathrm{tot})'=\pm \frac{1}{2}$. (a) For $S = 1$, the selection rules permit four possible tunneling excitations from the (approximately) doubly-screened ground state (blue, $F\simeq 2$), three from the singly screened (black), while only two survive when the ground state is unscreened (red, $F = 0$). 
	(b) $S = \frac{3}{2}$ shows corresponding behavior.
	Parameters: $K=100t$, $V=70t$, $J=-2t$, as indicated by gray dashed lines in Fig.\  \ref{fig:higher_spin_phase_diagrams_iso}(a) and (b).
	 }
	\label{fig_excitation_higherspins}
\end{figure*}

We can obtain additional insights into the phase boundaries by explicitly computing the RKKY energy $E_\mathrm{RKKY}$ in the various phases as a function of $S$. This also allows us to ask about the limit of large impurity spins, $S\rightarrow \infty$. In general, the RKKY coupling $J\mathbf{S}_1\cdot \mathbf{S}_2$ between spins of magnitude $S_1$ and $S_2$ coupling into a total spin of $S_\mathrm{tot}$ has the magnitude
\begin{equation}
    E_\mathrm{RKKY}= \frac{J}{2}[S_\mathrm{tot}(S_\mathrm{tot}+1)-S_1(S_1+1)-S_2(S_2+1)]. 
\end{equation}
Assuming, say, ferromagnetic coupling ($S_\mathrm{tot}=S_1+S_2$), one might then naively expect that the difference in RKKY energies of the fully unscreened phase ($S_1=S_2=S$) and the singly screened phase ($S_1=S$ and $S_2=S-\frac{1}{2}$) is $JS/2$ and thus proportional to $S$. A more careful treatment accounting for changes in the effective RKKY coupling $J$ between the different phases shows that the difference in RKKY energies approaches a constant independent of $S$ for $S\to\infty$. 

Given that the exchange coupling $K$ is large compared to the RKKY coupling, the projection theorem implies that the impurity spins $\mathbf{S}_1$ and $\mathbf{S}_2$ can be replaced by the effective screened spin of the adatom, $\mathbf{S}_\mathrm{eff}$,
albeit only up to an overall prefactor, which can be absorbed into a renormalization of the RKKY coupling $J$ 
(see App.\ \ref{app:higher_spin_dimer_iso_RKKY} for details). One finds that for each screened spin, the RKKY coupling is renormalized by a factor $1+1/(2S+1)$. Focusing first on ferromagnetic RKKY coupling, $J<0$, this gives the RKKY energies 
\begin{align}
\label{eq:FM RKKY energy higher spins}
    E_\mathrm{RKKY}=
    \begin{cases}
     JS^2 \phantom{ \left(\dfrac{1}{2S+1}\right) }  & S_{\textrm{tot}} = 2S,
     \\
     JS \left(S-\dfrac{1}{2S+1}\right)  &S_{\textrm{tot}} =2S-\frac{1}{2},
     \\ 
    J\left(S-\dfrac{1}{2S+1}\right)^2  & S_{\textrm{tot}} = 2S-1.
    \end{cases}
\end{align}
The phase boundary $E_\mathrm{YSR}=E_\mathrm{YSR}(J)$  between the unscreened ($S_\mathrm{tot}=2S$) and the singly screened ($S_\mathrm{tot}=2S-\frac{1}{2}$) phases follows from the energy balance 
\begin{equation}
    JS^2 = E_\mathrm{YSR} + JS \left(S-\dfrac{1}{2S+1}\right) - \tilde t. 
\end{equation}
Here, $\tilde t$ again denotes the energy gain of the singly screened phase due to the hybridization $t$ of the YSR states. Thus, we find 
\begin{equation}
    E_\mathrm{YSR} = \tilde t + \frac{JS}{2S+1}
    \label{eq:upperpb}
\end{equation}
for the phase boundary on the ferromagnetic side $J<0$. An analogous consideration for the phase boundary between the singly screened phase and the fully screened ($S_\mathrm{tot}=2S-1$) phase yields
\begin{equation}
 E_\mathrm{YSR} = -\tilde t + \frac{JS}{2S+1}
 - \frac{J}{(2S+1)^2}.
 \label{eq:lowerpb}
\end{equation}
As advertized above, the slope of these phase boundaries $E_\mathrm{YSR}=E_\mathrm{YSR}(J)$ approaches  $\frac{1}{2}$ in the limit of $S\to \infty$, which is distinctly different from the purely classical result of zero slope. Thus, the distinction between classical and quantum spins persists to arbitrarily large spins. 

The phase boundaries in Eqs.\ (\ref{eq:upperpb}) and (\ref{eq:lowerpb}) also imply that the singly screened phase terminates at $J=-2\tilde t (2S+1)^2$. 
For stronger ferromagnetic RKKY coupling $J$, there is a direct transition between the unscreened and the doubly screened phase with
\begin{equation}
 E_\mathrm{YSR} =   \frac{JS}{2S+1}
 \left( 1 - \frac{1}{2S(2S+1)}\right).
 \label{eq:directpb}
\end{equation}
describing the phase boundary. 

Antiferromagnetic RKKY interaction couples the impurity spins to $S_\mathrm{tot}=0$ (see App.\ \ref{app:higher_spin_dimer_iso_phase} for an exception). At large and positive $E_\mathrm{YSR}$, this results from a coupling of the unscreened impurity spins. As $E_\mathrm{YSR}$ is reduced and becomes negative, this eventually crosses over to coupling of the screened impurity spins. As for spin-$\frac{1}{2}$ dimers, this is a crossover rather than a phase boundary. We can deduce the crossover line by equating the energies of the $S_\mathrm{tot}=0$ states for unscreened and screened spins,
\begin{equation}
    -JS(S+1) = 2E_\mathrm{YSR}-2J\frac{(S-\frac{1}{2})(S+1)^2}{2S+1},
\end{equation}
which yields
\begin{equation}\label{eq:crossover_iso}
    E_{\textrm{YSR}}  = - J \frac{S + 1}{4S + 2}.
\end{equation}
Thus, the slope of the crossover line for antiferromagnetic RKKY coupling approaches $-\frac{1}{4}$ for large $S$.

We note that the zero-bandwidth model not only provides detailed insight into the physics of YSR dimers, but that the corresponding phase diagram for the spin-$1$ dimer in Fig.\ \ref{fig:higher_spin_phase_diagrams_iso}(a) is also in excellent qualitative agreement with full NRG calculations in Ref.\ \cite{Zitko2011}. 

\begin{figure}[t!]
         \centering
     \includegraphics[width=0.45\textwidth]{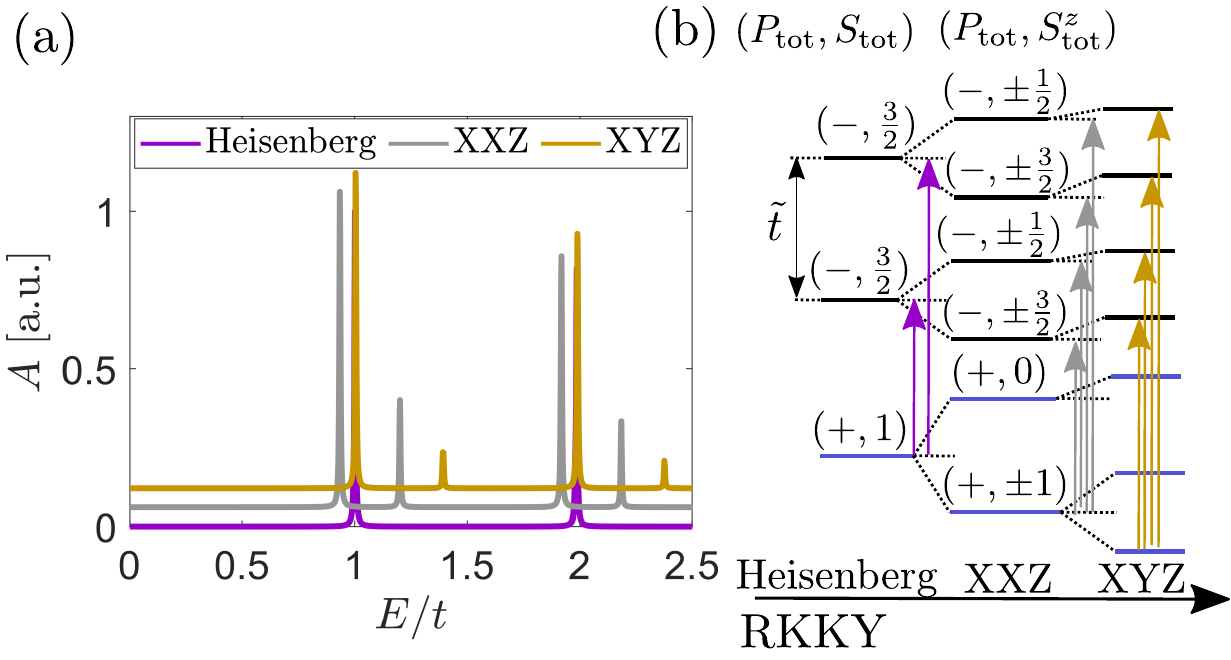}
	\caption{Tunneling spectroscopy of spin-$1$ dimers in the doubly screened region with isotropic exchange and ferromagnetic RKKY coupling. (a) Spectral functions for different symmetries of the RKKY coupling (offset for clarity). For Heisenberg RKKY coupling (purple), the resonances originate from tunneling into the symmetric and antisymmetric quartet states $(+,\frac{3}{2})$. Tunneling into  the symmetric and antisymmetric doublet states $(-,\frac{1}{2})$ is allowed, but beyond the energy range shown here (see Fig.\ \ref{fig_excitation_higherspins}).
	Upon breaking spin rotation symmetry, the symmetric and antisymmetric  quartets split into two peaks each. The odd-parity states remain degenerate even for XYZ coupling as they are protected by time reversal symmetry. (b) Corresponding level  schemes (not to scale). Parameters: $E_\mathrm{YSR}=-t$, $\Delta=10^6t$, $V = 2\Delta$, Heisenberg: $\hat{J}=-\mathrm{diag}(t,t,t)/\sqrt{3}$, XXZ: $\hat{J}=-\,\mathrm{diag}(t/2,t/2,t/\sqrt{2})$, XYZ:  $\hat{J}=-\,\mathrm{diag}(0.4t,0.58t,t/\sqrt{2})$.
	}
	\label{fig_spectral_fun_spin_1}
\end{figure} 

In Fig.\ \ref{fig_excitation_higherspins}, we show excitation spectra for adatom spins $S=1$ and $S=\frac{3}{2}$ along the dashed lines shown in Fig.\ \ref{fig:higher_spin_phase_diagrams_iso}. In particular, these spectra identify the transitions that are observable in tunneling experiments (see purple arrows in the figure). Single-electron tunneling flips the fermion parity and changes the total spin by $\pm 1/2$. For spin-$\frac{1}{2}$ dimers, the number of resonances of the tunneling spectra depends only on the fermion parity (Fig.\ \ref{fig_excitation12}). For higher spins, we find that it also varies with the total spin. If the ground state of the dimer has minimal ($S_\mathrm{tot}=0$) or maximal ($S_\mathrm{tot}=2S$) total spin, it is only coupled to the symmetric and antisymmetric states with $S_\mathrm{tot}=\frac{1}{2}$ or $S_\mathrm{tot}=2S-\frac{1}{2}$, respectively. This results in two tunneling resonances. If the total spin of the ground state differs from the minimal or maximal spin, as for ferromagnetic coupling of screened spins or in the half-integer-spin phases, tunneling couples to additional states and thus leads to further resonances as shown in Fig.\ \ref{fig_excitation_higherspins}. Corresponding spectral functions and spectra are illustrated in Fig.\ \ref{fig_spectral_fun_spin_1}.

\subsection{Single-ion anisotropy}
\label{subsec_anisotropy}

\begin{figure*}[t!]
    \centering
         \includegraphics[width=\textwidth]{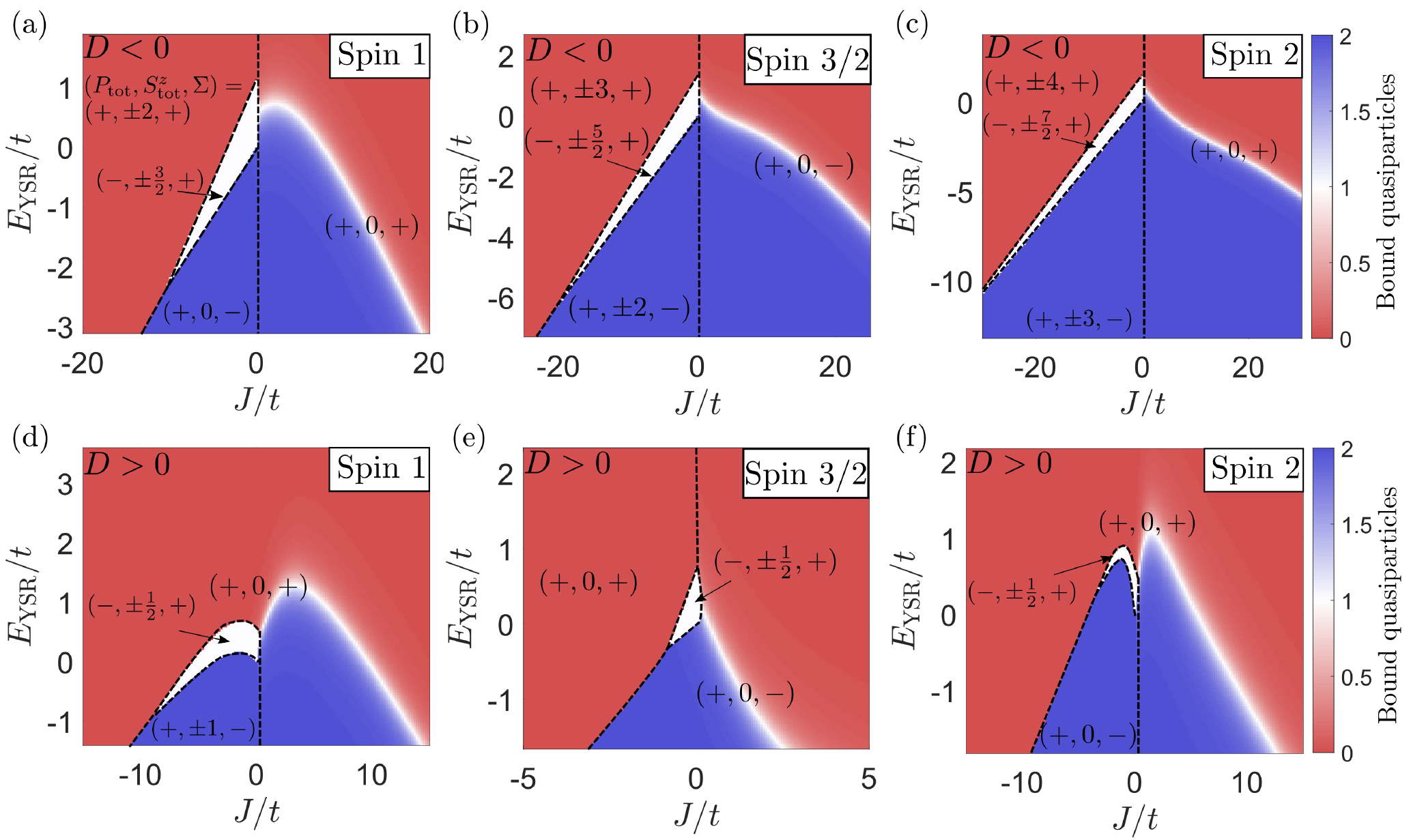}
	\caption{Phase diagrams for YSR dimers with higher spins $S$ (see panels) including  single-ion anisotropy. The phases are labeled by $(P_\mathrm{tot},S^z_{\mathrm{tot}},\Sigma)$, with $\Sigma$ indicating the spatial parity of the dimer. Note that $E_\mathrm{YSR}=E_o-E_e$ contains single-ion anisotropy, as defined in App.\ \ref{app:higher_spin_dimer_iso_mono}. (a)-(c) Easy-axis anisotropy $D<0$. For ferromagnetic RKKY coupling, the phase diagrams are unchanged compared to  Fig.~\ref{fig:higher_spin_phase_diagrams_iso} for isotropic couplings. For antiferromagnetic RKKY coupling, the phase boundaries are affected for $J\lesssim |D|$. (d)-(e) Easy-plane anisotropy $D>0$. The phase diagrams are strongly modified for $|J|\lesssim D$ with significant difference between integer and half-integer spins. For integer spins, weak RKKY coupling favors the screened phase, while for half-integer spins it strongly favors the unscreened phase. Integer spins have different spatial parity for the doubly screened phase, but equal parity in the unscreened phase. For half-integer spins, the roles are reversed. 
	Parameters: $V=2\Delta$, $\Delta = 10^6 t$, (a)-(c) $D=-10t$ , (d)-(f) $D=5t$.}
	\label{fig:higher_spin_phase_diagrams_aniso}
\end{figure*}

For strictly isotropic spin interactions, the ground and excited states are degenerate multiplets associated with different spin projections of $\mathbf{S}_\mathrm{tot}$. These degeneracies are lifted by anisotropic couplings as illustrated in Fig.\ \ref{fig_spectral_fun_spin_1}(a) and (b). This leads to splitting of tunneling resonances and affects phase diagrams. We illustrate the effects of magnetic anisotropy by studying uniaxial single-ion anisotropy,
\begin{align}
\label{eq_mag_anisotropy}
    H_\mathrm{ani}=D\sum\limits_j (S^z_j)^2.
\end{align}
For simplicity, we retain fully isotropic exchange coupling $K$ and RKKY interaction $J$. The results are only weakly affected by moderate anisotropy of $\hat{K}$ and $\hat{J}$, in particular if the type of anisotropy is consistent with the sign of $D$.

\subsubsection{Easy-axis anisotropy}

For $D < 0$ (easy-axis anisotropy), $H_\mathrm{ani}$ favors the maximal spin projections $S_\mathrm{eff}^z = \pm S$ (unscreened monomer) and  $S_\mathrm{eff}^z = \pm(S - 1/2)$ (screened monomer) of the monomer spins. Large and negative $D$ frustrates the transverse part of the RKKY interaction, so that the monomers act as effective Ising spins. However, it is important to note that the magnitude of these effective Ising degrees of freedom depends on the screening state of the adatom spin. For this reason, easy-axis anisotropy does not induce classical behavior of the dimer. (Classical behavior requires longitudinal exchange coupling $K$ in addition.)

Figure \ref{fig:higher_spin_phase_diagrams_aniso}(a)-(c) shows corresponding phase diagrams, which exhibit four distinct phases as for the fully isotropic model. The YSR energy $E_\mathrm{YSR}$ is now taken to include the single-anisotropy, see App.\ \ref{app:higher_spin_dimer_iso_mono} for details. We observe that the phase diagrams remain essentially unaffected by the single-ion anisotropy $D$ as long as the RKKY coupling is ferromagnetic. In this case, the easy-axis anisotropy merely selects the  states with maximal spin projection $S^z_{\textrm{tot}}$ from the degenerate ground-state multiplets for isotropic couplings. Thus, the unscreened phase has $S^z_{\textrm{tot}} = \pm 2S$, the singly screened phase has $S^z_{\textrm{tot}} = \pm (2S - 1/2)$, and the doubly screened phase has $S^z_{\textrm{tot}} = \pm (2S-1)$. 

In contrast, easy-axis anisotropy changes the phase diagrams qualitatively for antiferromagnetic RKKY coupling. We observe a change in slope of the crossover line, when the anisotropy and RKKY energies become comparable. For large RKKY coupling, $J\gg \abs{D}$, the anisotropy is a small perturbation. Then, the ground state has $S_{\textrm{tot}} = 0$ and the crossover line is given by Eq.\ \eqref{eq:crossover_iso} obtained in the absence of the anisotropy. In the opposite limit of small RKKY coupling, $J \ll \abs{D}$, the monomers behave like Ising degrees of freedom, which minimize the total spin projection by anti-aligning, $S^z_{\textrm{tot}} = 0$. The crossover line in this regime follows from the energy balance 
\begin{equation}
    -JS^2 = 2E_\mathrm{YSR} - J\left(1+\frac{1}{2S+1}\right)^2 \left(S-\frac{1}{2}\right)^2,
\end{equation}
where the last term on the right-hand side accounts for the renormalization of the RKKY coupling for screened spins. This yields a crossover line
\begin{equation}
    E_\mathrm{YSR} = -\frac{JS}{2S+1}\left(1-\frac{1}{2S(2S+1)}\right),
\end{equation}
with a steeper slope than at large $J$. 

The monomer spin of $S=1$ is a special case. While the unscreened monomers act like Ising spins, the screened monomers are effective spin-$\frac{1}{2}$ degrees of freedom and gain energy through the transverse part of the RKKY interaction regardless of the magnitude of $J$. For antiferromagnetic RKKY interaction, the screened monomer spins couple into a singlet. Accounting for the renormalization of the RKKY coupling for screened spins, the gain in RKKY energy due to singlet formation is  $E_{\textrm{RKKY}} = -4J/3$. Thus, for weak RKKY coupling the crossover line follows from the energy balance
\begin{equation}
    -J = 2E_\mathrm{YSR} - \frac{4J}{3},
\end{equation}
which predicts a crossover line
\begin{equation}
    E_\mathrm{YSR} = \frac{J}{6}
\end{equation}
with a positive slope. Interestingly, in this regime, small RKKY coupling favors the screened over the unscreened state. We finally observe that surprisingly, ferromagnetic RKKY interaction couples the screened monomers into the triplet state with  $S^z_{\textrm{tot}} = 0$. The screened monomers have $S_\mathrm{eff}=\frac{1}{2}$ and the easy-axis anisotropy does not affect them directly. However, unlike the $S^z_{\textrm{tot}} = \pm 1$ states, the $S^z_{\textrm{tot}} = 0$ state gains energy by admixing the antialigned state of the unscreened dimer.

\subsubsection{Easy-plane anisotropy}

The phase diagrams for easy-plane anisotropy in Fig.\ \ref{fig:higher_spin_phase_diagrams_aniso}(d)-(f) show more substantial differences compared to fully isotropic couplings. First, there are qualitative differences between integer and half-integer $S$. For half-integer spins, RKKY coupling favors the unscreened state as for isotropic couplings. In contrast, for integer spins, RKKY coupling initially favors the screened over the unscreened states and the conventional behavior is recovered only once the RKKY coupling $J$ exceeds the anisotropy $D$ in magnitude. Second, we find that unlike in the isotropic and easy-axis cases, the phase diagrams display only three distinct phases. This results from the different behavior of the phase boundaries at $J=0$, which separates ferromagnetic and antiferromagnetic phases. For easy-plane anisotropy, the extent of this phase boundary depends on the nature of the adatom spins $S$. For integer adatom spins, there is no such phase boundary for unscreened adatom spins, while for half-integer adatom spins, the phase boundary is absent for screened spins. 

The different behaviors of integer and half-integer monomer spins $S$ at weak RKKY coupling, $\abs{J}\ll \abs{D}$, can be understood as follows. The easy-plane anisotropy favors small spin projections of the monomers. In the unscreened state, anisotropy favors $S_\mathrm{eff}^z=0$ for integer $S$ and $S^z_\mathrm{eff}=\pm \frac{1}{2}$ for half-integer spins. The situation is reversed in the screened state. When the single-ion anisotropy favors $S_\mathrm{eff}^z=\pm \frac{1}{2}$, the adatom acts effectively as a spin-$\frac{1}{2}$ degree of freedom. 

We first consider integer-spin monomers. In the unscreened state, the monomers are in $S^z=0$ states and the RKKY coupling is ineffective. Correspondingly, there is a single $S^z_\mathrm{tot}=0$ phase regardless of the sign of the RKKY coupling. In contrast, the screened monomers effectively act as spin-$\frac{1}{2}$ degree of freedom. The effective RKKY coupling written in terms of these spin-$\frac{1}{2}$ degrees of freedom has an easy-plane anisotropy ($S=1$ monomers are again an exception), so that the ground state has $S^z_\mathrm{tot}=0$ even for ferromagnetic RKKY interaction. Importantly, however, there is still a phase boundary at $J=0$. While the phases are not distinguished by fermion parity or spin projection, the phases differ in their spatial parity $\Sigma$ as defined in Eq.\ (\ref{eq:SigmaDef}). Under spatial parity, the spin ground state is symmetric for ferromagnetic RKKY coupling and antisymmetric for antiferromagnetic coupling. As $\Sigma$ also interchanges the fermions, it further distinguishes the unscreened and doubly-screened phases in the case of ferromagnetic RKKY coupling. Figure \ref{fig:higher_spin_phase_diagrams_aniso} explicitly displays $\Sigma$ for all phases. 

For half-integer spins, the behavior of unscreened and screened monomers are essentially reversed. This explains the difference in the phase boundaries at $J=0$. The difference in the sign of the slope of the phase boundary at weak RKKY coupling can be understood as follows. There is a gain in RKKY energy only when there is a residual spin-$\frac{1}{2}$ degree of freedom. For integer spin, this is the case for screened monomers and consequently, RKKY coupling favors the screened phases. In contrast, it is the unscreened monomers which retain a residual spin-$\frac{1}{2}$ degree of freedom for half-integer spins and the RKKY coupling favors the unscreened phases. 

We finally note that strong RKKY coupling, $\abs{J}\gg \abs{D}$, couples the monomer spins into a total spin $\mathbf{S}_\mathrm{tot}$. In view of the projection theorem, the monomer spins are effectively proportional to $\mathbf{S}_\mathrm{tot}$, which is integer for both the unscreened and the fully screened phases. Then,
the single-ion anisotropy favors the $S^z_\mathrm{tot}=0$ state regardless of the sign of the RKKY coupling. 

\subsection{Anisotropy and Dzyaloshinsky-Moriya interactions}
\label{subsec:anisotropic couplings}

\begin{figure}[t!]
         \centering
         \includegraphics[width=0.4\textwidth]{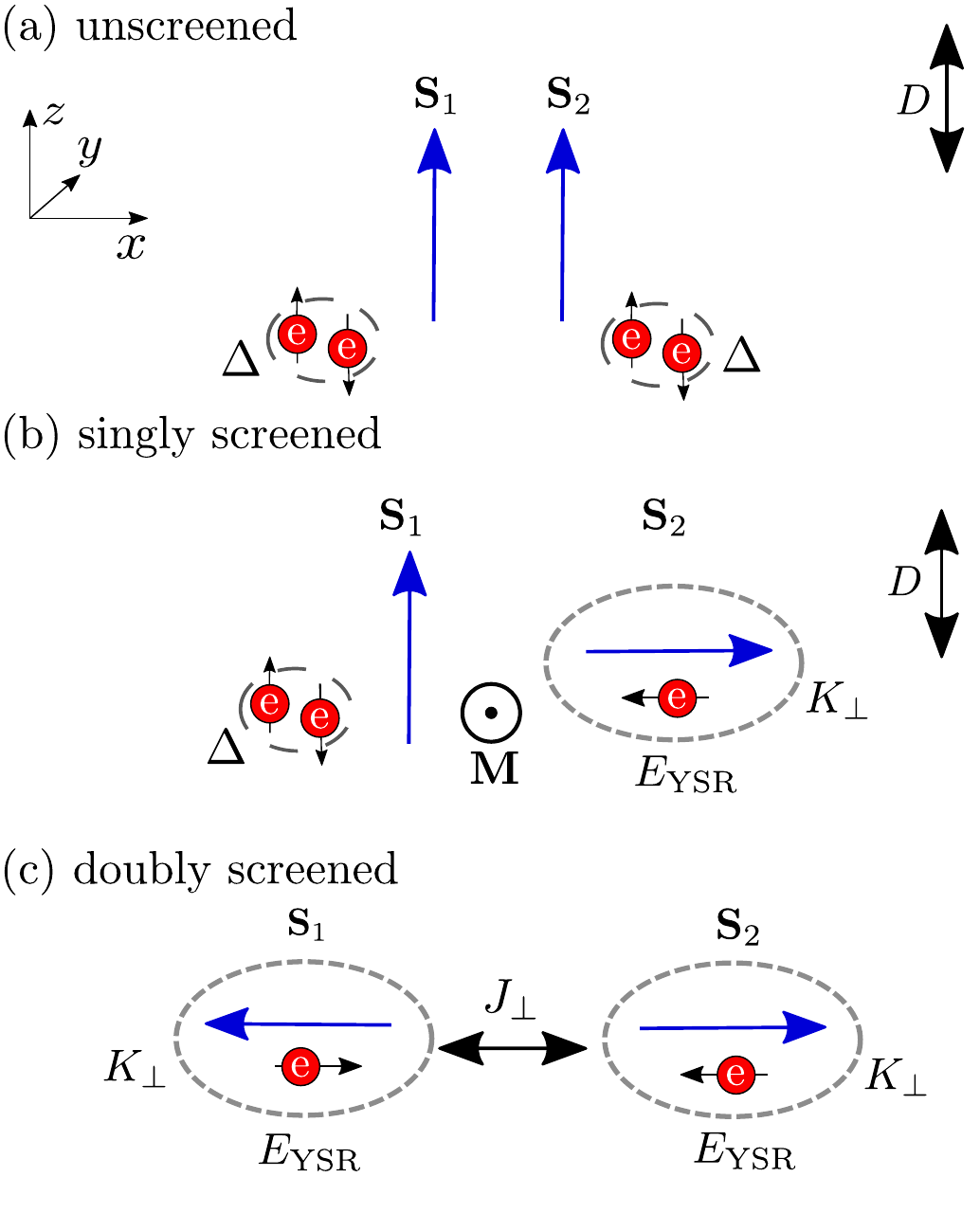}
         \caption{Classical spin alignment for different screening scenarios in the presence of anisotropic coupling, easy-axis single-ion anisotropy,  and DM interaction, as outlined in Sec.~\ref{subsec:anisotropic couplings}. (a) Unscreened spins $\mathbf{S}_1$ and $\mathbf{S}_2$  align along the $z$-direction due to easy-axis single-ion anisotropy $D$. (b) A screened spin (here $\mathbf{S}_2$) aligns perpendicular to the $z$-direction due to transverse exchange coupling $K_\perp$, which dominates over the single-ion anisotropy. The singly screened case is favored by DM interactions with $\mathbf{M}$ lying in the $xy$-plane. (c) Two screened spins align in the $xy$ plane (here antiferromagnetically), associated with a gain in energy due to the transverse RKKY coupling. Quantum effects modify the classical picture, see Fig.~\ref{fig_Yazdani}.}
         \label{fig_yazdani_sketch}
\end{figure}

\begin{figure*}[t!]
         \centering
         \includegraphics[width=\textwidth]{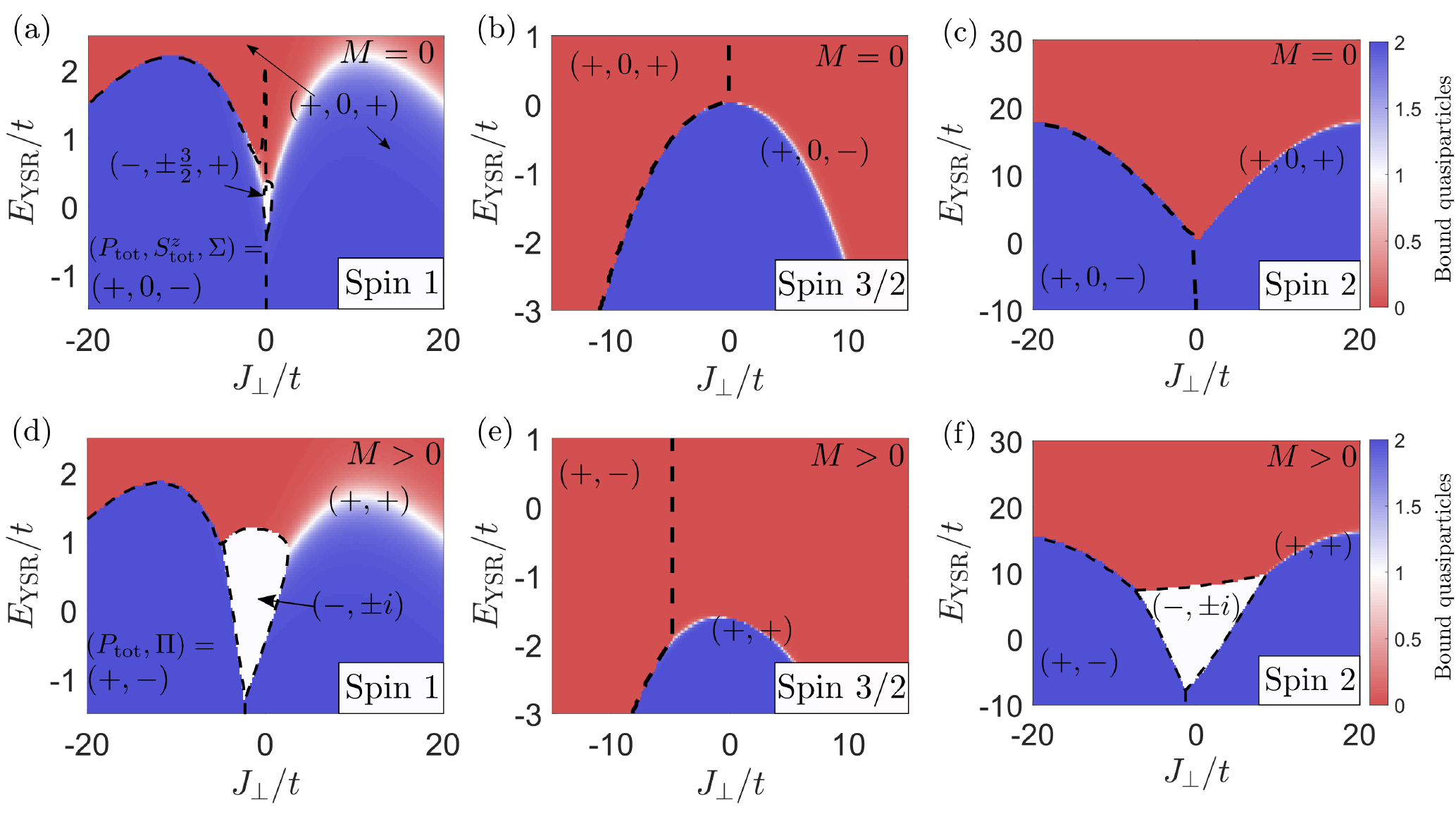}
     \caption{ Phase diagrams for YSR dimers with transverse ($xy$) couplings  $K_\perp$ and $J_\perp$, easy-axis anisotropy $D$, and DM interaction $M$. 
        (a)-(c) For vanishing DM interaction, the fermion parity $P_\mathrm{tot}$, the spin projection $S^z_\mathrm{tot}$, and the spatial parity $\Sigma$ are good quantum numbers. For integer spin, small RKKY coupling $J_\perp \ll |D|$ favors the screened spin phases as expected from classical considerations, see Fig.\ \ref{fig_yazdani_sketch}. In contrast, for half-integer spins the screened phase is suppressed. Note also the alternation in phase boundaries associated with changes in spatial parity $\Sigma$. The singly-screened phase (white) is increasingly suppressed for higher spins due to the different spin alignments of unscreened and screened monomers, see Fig.\ \ref{fig_yazdani_sketch}. 
     (d)-(e) For nonzero DM interaction with $\mathbf{M}=M \hat{\mathbf{y}}$ orthogonal to the single-ion anisotropy, only $P_\mathrm{tot}$ and the discrete spin rotation $\Pi$ remain good symmetries. DM interaction drastically enhances the singly-screened phase (white) for integer spins, while it merely shifts the phase boundary/crossover for half-integer spins.  Parameters: $V=2\Delta$, $\Delta = 10^6 t$, $D=-15t$, (a)-(c) $M=0$, (d)-(f) $M=5t$.}
	\label{fig_Yazdani}
\end{figure*} 

Anisotropy of the exchange coupling $K$ strongly affects the phase diagram. As we have seen above, purely longitudinal (Ising-like) coupling with $K_\perp=0$ effectively corresponds to a classical-spin model, leading to qualitatively distinct phase diagrams from quantum spins. For spin-$\frac{1}{2}$ adatoms, the classical spin model remains adequate only as long as the transverse exchange coupling $K_\perp$ is small compared to the hybridization and RKKY coupling between the adatom spins. 
As we show in App.\ \ref{app:higher_spin_dimer_exchange_aniso}, for higher spins and ferromagnetic RKKY coupling, the condition for classical behavior is $K_\perp\ll K_z$. For antiferromagnetic RKKY coupling, $J_\perp\ll J_z$ has to be satisfied, in addition. 
Since anisotropy of $\hat K$ is a consequence of spin-orbit coupling and diminishes under Kondo scaling, this is a rather stringent condition. Note that the zero-bandwidth model can be viewed as the result of integrating out the quasiparticle continuum, so that its exchange coupling should effectively account for Kondo renormalizations down to scales of order $\Delta$. 

For $xy$-like exchange coupling, i.e., when $K_\perp$ dominates over $K_z$, the exchange coupling forces the effective adatom spin $\mathbf{S}_\mathrm{eff}$ of the screened state into the $xy$-plane. When this occurs along with easy-axis single-ion anisotropy, screening of the adatom spin is accompanied by a reduction of $S^z_\mathrm{eff}$ from its maximal value ($S^z_\mathrm{eff}=\pm S$) in the unscreened state to its minimal value ($S^z_\mathrm{eff}=0$ or $\pm \frac{1}{2}$)
in the screened state (see Fig.\ \ref{fig_yazdani_sketch} for an illustration). It was recently argued \cite{Ding2021} that this  situation is realized for gadolinium adatoms ($S=7/2$) on a bismuth surface with proximity-induced superconductivity. Moreover, it was argued that the dimer spins interact via RKKY coupling dominated by $J_\perp$ as well as Dzyaloshinsky-Moriya coupling.

Corresponding phase diagrams (with $K_z=J_z=0$) are shown in Fig.\ \ref{fig_Yazdani}. First consider Figs.\ \ref{fig_Yazdani}(a)-(c) for vanishing DM coupling. Interestingly, the transverse RKKY coupling strongly favors the screened states for integer spins, with the phase boundary being cusp-like at $J_\perp=0$. In contrast, RKKY coupling still favors the unscreened state for half-integer spins, with a smooth phase boundary at $J_{\perp}=0$. 

This stark difference between integer and half-integer spins can be understood as follows. For integer adatom spin, the effective spin of the screened monomer is half-integer, so that the transverse exchange coupling $K_\perp$ favors $S_\mathrm{eff}^z=\pm \frac{1}{2}$ and the adatom acts as an effective spin-$\frac{1}{2}$ degree of freedom. Transverse RKKY interaction of either sign couples these effective spins into $S_\mathrm{tot}^z=0$ states already at linear order in $J_\perp$. In contrast, the transverse RKKY coupling couples the ground state of the unscreened dimer only to excited states, so that the RKKY energy is quadratic in $J_\perp$. As a result, the gain in RKKY energy is parametrically larger for and thus favors the doubly-screened dimer.  

For half-integer spins, the situation remains unchanged for the unscreened dimer. In contrast, the screened monomers are now in a $S_\mathrm{eff}^z=0$ state. As a result, the transverse RKKY coupling couples only to excited states of the monomer, and the gain in RKKY energy is also quadratic in $J_\perp$, leading to a smooth boundary between unscreened and doubly-screened regions. The curvature of the boundary at small $J_\perp$ can be deduced by noting that the excited states lead to energy denominators of order $D$ for the unscreened monomer and of order $K_\perp$ for the screened monomer. This implies a larger gain in RKKY energy for the unscreened dimer, so that the phase boundary in Fig.\ \ref{fig_Yazdani}(b) bends downward \footnote{We note that the energy balance inverts, when $K_\perp \lesssim \abs{D}S^4$. Corresponding phase diagrams are included and discussed in App.\ \ref{app:higher_spin_dimer_iso_dmi}.}. Some of the phase boundaries are again associated with changes in the spatial parity $\Sigma$. For integer spins, the spatial parity of the doubly-screened phases depends on the sign of $J_\perp$ and there is a phase boundary at $J_\perp=0$. There is no corresponding phase boundary for the unscreened dimer. The roles of screened and unscreened phases are reversed for half-integer spins. 

It is also evident in Figs.\ \ref{fig_Yazdani}(a)-(c) that the extent of the singly-screened phase is strongly reduced as the adatom spin $S$ increases. In the present situation, the projections of the effective spins of unscreened and screened monomer differ by more than $\frac{1}{2}$ (spin-$1$ monomers are an exception). Since the intermonomer hopping $t$ transfers only spin $\frac{1}{2}$, hybridization splittings require increasingly high-order processes as the monomer spin $S$ increases. For the same reason, the crossover between molecular and local singlets for antiferromagnetic RKKY coupling becomes sharper with increasing $S$. 

As shown in Figs.\ \ref{fig_Yazdani}(d)-(f), a robust partially screened phase appears for integer spins once we include Dzyaloshinsky-Moriya coupling  
\begin{align}\label{eq:dmi_IA}
    H_\mathrm{DM} = \mathbf{M}\cdot (\mathbf{S}_1 \times \mathbf{S}_2)
\end{align}
with $\mathbf{M}=M\mathbf{\hat y}$ lying in the $xy$-plane. DM of this form was argued to be relevant in the above-mentioned experiment by Ding et al.\ \cite{Ding2021}. Evidently, this DM coupling leads to a substantial gain in energy of the singly-screened phase, where the spins of the two monomers are effectively at right angles, but no corresponding gains in the unscreened and doubly screened regions, where the monomer spins  effectively align in parallel. As a result, the partially-screened phase can become the ground state over an extended parameter range, showing that DM coupling provides an alternative mechanism for the emergence of a singly-screened phase. While this semiclassical picture rationalizes the results for integer spins, a more quantum mechanical understanding of the emergence of the singly-screened phase is needed for half-integer spins, see App.\ \ref{app:higher_spin_dimer_iso_dmi}. 

In the presence of DM interactions, neither the spin projection $S_\mathrm{tot}^z$ nor the spatial parity $\Sigma$ are conserved quantum numbers. There is, however, a discrete spin rotation symmetry 
\begin{align}
\Pi =\exp(i \pi S_\mathrm{tot}^y),
\end{align}
which leaves both the single-ion anisotropy and the DM interaction invariant. For $P_\mathrm{tot}=1$, the dimer is an integer spin system and  $\Pi^2=1$, so that $\Pi$ has eigenvalues $\pm 1$. For $P_\mathrm{tot}=-1$, in contrast,  the dimer has half-integer spin and $\Pi^2=-1$, so that $\Pi$ has eigenvalues $\pm i$. Figures \ref{fig_Yazdani}(d)-(f) label the phases by the quantum numbers $P_\mathrm{tot}$ and $\Pi$. 

It is interesting to distinguish more explicitly between classical and quantum aspects of the phase diagrams in Fig.\ \ref{fig_Yazdani}. In a classical treatment, one minimizes the energy as a function of the classical spin configuration. In the absence of the DM interaction and the hybridization $t$, the unscreened dimer has energy 
\begin{equation}
    E_{ee} = -2\abs{D}S^2 \cos^2\alpha -\abs{J_\perp} S^2 \sin^2\alpha,
\end{equation}
where $\alpha$ is the angle of the adatom spins relative to the $z$-axis. Minimizing with respect to $\alpha$ predicts an abrupt transition between spin polarizations along the $z$-axis at small $\abs{J_\perp}$ and in the $xy$-plane at large $\abs{J_\perp}$. In contrast, the doubly-screened dimer has spins that lie in the $xy$ plane regardless of $J_\perp$, so that its energy equals 
\begin{equation}
    E_{oo} = 2E_\mathrm{YSR} -\abs{J_\perp} S^2.
\end{equation}
Then the phase boundary follows from $E_{oo}=E_{ee}$, which yields
\begin{equation}
    E_\mathrm{YSR} = \frac{1}{2}(-2\abs{D}  +\abs{J_\perp} )S^2\cos^2\alpha,
\end{equation}
with an abrupt transition between $\cos\alpha=1$ at small $J_\perp$ and $\cos\alpha=0$ at large $J_\perp$. This is consistent with the initial cusp-like behavior of the phase boundaries at small $J_\perp$, which we find for integer adatom spins in Fig.\ \ref{fig_Yazdani}(a) and (c). However, the phase boundary of the classical phase diagram saturates to a $J_\perp$-independent value of $E_\mathrm{YSR}$ for $J_\perp$ beyond the abrupt transition, while the phase boundary bends downward at large $J_\perp$ in the quantum phase diagram. This bending is therefore a quantum effect, originating in the reduction of the effective spin by screening. Moreover, the alternation in the phase diagrams between integer and half-integer spins is specific to quantum spins. Interestingly, the phase diagram for half-integer spins is distinctly different from the classical phase diagram.  
\section{Conclusions} 
\label{sec:con}

Motivated by recent experiments, we have discussed Yu-Shiba-Rusinov dimers of quantum spins. Both Kondo resonances and spin excitations of individual adatom spins show that magnetic adatoms on superconductors act as quantum spins. This contrasts with theoretical discussions of YSR dimers, which focus on a description in terms of classical spins. Moreover, the existing works on quantum YSR dimers are limited to small adatom spins and do not account for single-ion anisotropy or DM interactions, which are relevant for the transition-metal and rare-earth systems used in many of the experiments. 

At first sight, one may expect that the physics of quantum YSR dimers becomes more classical for larger impurity spins or stronger (easy-axis) single-ion anisotropy. Remarkably, we find that this is not the case and the qualitative differences between the phase diagrams for classical and quantum spins persist even in the limit of $S\to\infty$. These differences are rooted in the distinct screening properties of classical and quantum spins. While in both cases, binding of a quasiparticle induces a quantum phase transition with increasing exchange coupling between adatom spin and substrate electrons, this transition is associated with Kondo-like screening for quantum spins only. 

The screening of the adatom spins by bound quasiparticles directly modifies the RKKY energy of the dimer. Since the RKKY energy is typically larger for higher adatom spins, one frequently finds that RKKY coupling stabilizes the unscreened phases. In some cases, however, the RKKY coupling can also stabilize the screened phases, for instance when easy-plane single-ion anisotropy frustrates the RKKY coupling of the unscreened phases. This example also shows that even such gross features of the phase diagrams can depend on whether the adatoms have integer or half-integer spins. 

In the present paper, we have focused on illustrating basic phenomena in quantum YSR dimers. Evidently, this system has an exceedingly rich parameter space, and several aspects are left for future work. For instance, we have restricted attention to a single conduction-electron channel per magnetic adatom. More generally, higher spins couple to multiple conduction electron channels, admitting for multistage screening of the adatom spins. While coupling to multiple conduction electron channels can be addressed within the zero-bandwidth model employed here, there are also interesting aspects which are beyond the reach of this approach. Most notably, this concerns the spatial structure of the hybridized YSR wave functions which are readily resolved in state-of-the-art experiments. 

Recent works have made proposals for qubits based on YSR dimers \cite{Mishra2021,Pavesic2021}. It seems likely that a thorough theoretical understanding of such qubits would have to rely on a description of the magnetic impurities in terms of quantum spins. Moreover, possible implementations in transition-metal or rare-earth systems require one to account for magnetic anisotropy or noncollinear spin coupling as we have done here.

\begin{acknowledgments}
We gratefully acknowledge funding by Deutsche Forschungsgemeinschaft through CRC 183 (project C03) as well as CRC 910 (project A11). 
\end{acknowledgments}

\appendix

\section{Spin-$\frac{1}{2}$ adatoms}
\label{app:spin12-old}

This appendix discusses the zero-bandwidth model (see also \cite{Oppen2021,Steiner2021}) for spin-$\frac{1}{2}$ adatoms, see Eq.\  \eqref{eq:monomodel} for the monomer and Eq.\ \eqref{eq:model12} for the dimer. 

\subsection{Spin-$\frac{1}{2}$ monomer}
\label{app:spin12_monomer}

The lowest-energy state for even fermion parity is a direct product of a free impurity spin $\ket{\Uparrow/\Downarrow}$ and the paired BCS state $\ket{\textrm{BCS}}  = (u+vc^\dagger_{ \downarrow} c^{\dagger}_{ \uparrow}) \ket{0}$ with amplitudes 
\begin{subequations}
   \begin{align}
    u^2=&\ \frac{1}{2}\pqty{1+\frac{V}{\sqrt{\Delta^2+V^2}}},\\
    v^2=&\ \frac{1}{2}\pqty{1-\frac{V}{\sqrt{\Delta^2+V^2}}}.
\end{align} 
\end{subequations}
This state has energy 
\begin{equation}
    E_e = V - \sqrt{\Delta^2 +V^2} 
    \label{eq:Eeven12}
\end{equation}
independent of the exchange coupling $K$.

For odd fermion parity, the lowest-energy state binds a quasiparticle and has energy 
\begin{equation}
    E_{\textrm{o}} = V- \frac{1}{4}(K_z + 2 K_{\perp})
\end{equation}
independent of the pairing strength $\Delta$. For purely longitudinal (Ising-like) exchange coupling, the state is a doublet
$\ket{\Uparrow,\downarrow}$, $\ket{\Downarrow,\uparrow}$. For nonzero transverse coupling, there is a unique singlet ground state
\begin{equation}
    \ket{s} = \frac{1}{\sqrt{2}} \pqty{\ket{\Uparrow,\downarrow}-\ket{\Downarrow,\uparrow} },
\end{equation}
corresponding to a screened adatom spin. 

The quantum phase transition between even-fermion-parity (weak coupling) and odd-fermion-parity (strong coupling) ground states occurs when $E_o = E_e$. The energy of the subgap YSR excitation 
\begin{equation}
  E_\mathrm{YSR} = E_o-E_e  = \sqrt{\Delta^2 + V^2} - \frac{1}{4}\pqty{K_z + 2K_{\perp}}
\end{equation}
vanishes at the quantum phase transition.

\begin{figure*}[t!]
    \centering
    \includegraphics[width=\textwidth]{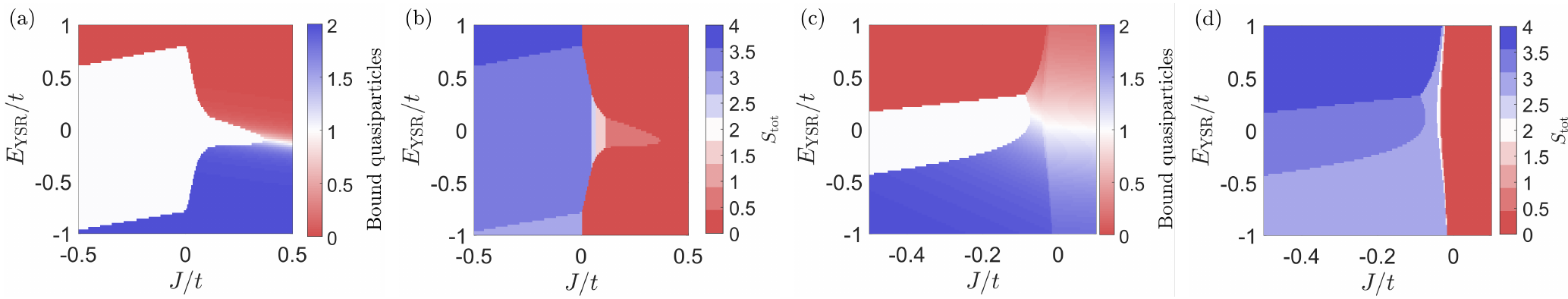}
    \caption{Phases of a $S=2$ YSR dimer with isotropic exchange coupling and weak isotropic RKKY interaction $J$. (a),(c) show number of bound quasiparticles $F$ and (b),(d) show total spin $S_\mathrm{tot}$ as a function of $J$ and $E_{\textrm{YSR}}$. Parameters: $\Delta = 10^6t$, (a),(b) $V=20\Delta$ and (c),(d) $V=0.5\Delta$.}
    \label{fig:higher_spin_phase_diagrams_iso_app}
\end{figure*}

\subsection{Spin-$\frac{1}{2}$ dimer}
\label{app:spin12_dimer_iso}

For isotropic couplings $K=K_z=K_\perp$ and $J=J_z=J_\perp$, states can be labeled by the total fermion parity $P_\mathrm{tot}$ as well as the total spin $S_\mathrm{tot}$ and its projection $S_\mathrm{tot}^z$. Below, we measure dimer energies relative to the energy of two uncoupled and unscreened monomers.   

For ferromagnetic coupling, the unscreened phase ($P_\mathrm{tot}=1, S_\mathrm{tot}=1$) has energy $-\frac{1}{4}J$ due to RKKY coupling of the adatom spins into a molecular triplet. The doubly-screened phase ($P_\mathrm{tot}=1, S_\mathrm{tot}=0$) does not gain RKKY energy and has energy $2E_\mathrm{YSR}$. These energies are unaffected by the hybridization $t$. Thus, the phase boundary at large and negative RKKY coupling follows $E_\mathrm{YSR}=J/8$.

For antiferromagnetic coupling, we first ignore the hybridization $t$. Then, the unscreened ground state ($P_\mathrm{tot}=1, S_\mathrm{tot}=0$) has energy $-\frac{3}{4}J$ due to RKKY coupling into a molecular singlet. The doubly-screened state has the same quantum numbers and energy $2E_\mathrm{YSR}$ independent of the RKKY coupling. This leads to a crossover line at $E_\mathrm{YSR}=-3J/8$. The width of the crossover is governed by the nonzero matrix element of the hybridization $t$ between these states, 
\begin{equation}
    \tilde{\Delta} = \sqrt{2} t u v , 
\end{equation}
which can be interpreted as an effective nonlocal singlet pairing \cite{Steiner2021}. In the presence of $t$, the ground state for antiferromagnetic RKKY coupling is thus a superposition of the local and molecular singlets with energy  
\begin{equation} 
\label{eq:12_isotropic_energy_stot=0}
    E_{\textrm{YSR}}   - \frac{3}{8} J - \sqrt{\tilde{\Delta}^2 + \pqty{E_{\textrm{YSR}}   + \frac{3}{8} J}^2}. 
\end{equation}

The odd-parity phase ($P_\mathrm{tot}=-1, S_\mathrm{tot}=\frac{1}{2}$) derives from the hybridization of the two configurations with a screened and an unscreened monomer into symmetric and antisymmetric superpositions. The corresponding splitting is equal to \cite{Steiner2021}
\begin{equation}
    2\tilde{t} = t\left(u^2-v^2\right). 
\end{equation}
Thus, the lower-energy superposition has energy $E_{\textrm{YSR}} - \tilde{t}$.

The expressions for the ground-state energies of the various phases can be used to find analytical results for the phase boundaries seen in Fig.\ \ref{fig_QuCl}(b). We also note that these analytical results can be readily extended to anisotropic exchange or RKKY coupling. The condition mentioned in Sec.\ \ref{sec_dimer_12} for the applicability of the classical phase diagram follows from such a calculation. 

\section{Higher-spin adatoms}
\label{app:higher_spin_dimer_iso}

\subsection{Monomers}
\label{app:higher_spin_dimer_iso_mono}

For isotropic exchange coupling and in the absence of uniaxial anisotropy, the lowest-energy state with even fermion parity has the same energy $E_e$ as for the spin-$\frac{1}{2}$ dimer, see Eq.\ \eqref{eq:Eeven12}. The lowest-energy state with odd fermion parity has energy 
\begin{equation}
    E_o = V-\frac{S+1}{2}K.
\end{equation}
This yields Eq.\ \eqref{eq:EYSR_iso} for the YSR energy $E_\mathrm{YSR}=E_o-E_e$.

For anisotropic exchange couplings $K_{\perp} \neq K_z$ or with uniaxial anisotropy $D$, fermion parity and $S^z_{\textrm{eff}}$ are conserved quantities. The low-energy states of the unscreened monomer have energy (assuming large $\Delta$ and $K$, as usual)
\begin{equation}
    E_e(S^z_{\textrm{eff}}) = V - \sqrt{\Delta^2 +V^2} +D (S^z_{\textrm{eff}})^2.
\end{equation}
For $D<0$, the lowest-energy state has maximal spin projection $S^z_{\textrm{eff}} = \pm S$, with energy $E_e=E_e(\pm S)$. For $D>0$, the lowest-energy state has minimal spin projection, i.e., $S^z_{\textrm{eff}} = 0$ for integer $S$ [lowest-energy state with energy $E_e=E_e(0)$] and $S^z_{\textrm{eff}} = \pm 1/2$ for half-integer $S$ [lowest-energy state with energy  $E_e=E_e(1/2)$]. 
For the screened monomer, the states are spanned by the basis $\ket{S,S^z}\otimes\ket{\sigma}$. By conservation of $S^z_{\textrm{eff}}$, the Hamiltonian couples only states $\ket*{S,S^z_{\textrm{eff}}-\frac{1}{2}}\otimes\ket{\uparrow}$ and $\ket*{S,S^z_{\textrm{eff}}+\frac{1}{2}}\otimes\ket{\downarrow}$ where $- S + 1/2 \leq S^z_{\textrm{eff}} \leq S - 1/2$. They are coupled by
\begin{multline}
    h(S^z_{\textrm{eff}}) = V - \frac{K_z}{4} + D \bqty{(S^z_{\textrm{eff}})^2+\frac{1}{4}} \\ + \frac{1}{2} \Big[ (K_z - 2D) S^z_{\textrm{eff}} \rho_3 + K_{\perp} \alpha(S^z_{\textrm{eff}}) \rho_1 \Big] ,
\end{multline}
where $\rho_i$  are Pauli matrices acting in this subspace and we define   $\alpha(S^z_{\textrm{eff}}) = \sqrt{S (S+1)-(S^z_{\textrm{eff}})^2 + 1/4}$. Diagonalizing $h(S^z_{\textrm{eff}})$ yields the low-energy eigenstates 
\begin{equation}
\ket{S^z_{\textrm{eff}}} =
    d_- \ket*{S,S^z_{\textrm{eff}} - \frac{1}{2}}\otimes\ket{\uparrow} - d_+ \ket*{S,S^z_{\textrm{eff}}+ \frac{1}{2}}\otimes\ket{\downarrow}, \label{eq: eigenstates screened}
\end{equation}
with energy 
\begin{align}
 &E_o(S^z_{\textrm{eff}}) = V - \frac{K_z}{4} + D \pqty{(S^z_{\textrm{eff}})^2+\frac{1}{4}} \nonumber \\ 
    &\quad\quad\quad - \frac{1}{2}\sqrt{(K_z -2D)^2 (S^z_{\textrm{eff}})^2 + K_{\perp}^2 \alpha^2(S^z_{\textrm{eff}})} \label{eq: spectrum screened impurity} 
\end{align}
and amplitudes
\begin{multline}
    d_{\pm}^2(S^z_{\textrm{eff}}) =    \frac{1}{2} \pm \frac{(K_z -2D)S^z_{\textrm{eff}}}{2\sqrt{(K_z -2D)^2 (S^z_{\textrm{eff}})^2 + K_{\perp}^2 \alpha^2(S^z_{\textrm{eff}}) }}. 
    \label{dplusminus}
\end{multline}
We ignore extremal $S^z_{\textrm{eff}}$, i.e., states $\ket{S,S}\otimes\ket{\uparrow}$ and $\ket{S,-S}\otimes\ket{\downarrow}$, as they do not contribute at low energy.
The lowest-energy state is $E_o=\min_{S^z_{\textrm{eff}}}E_o(S^z_{\textrm{eff}})$. We again define $E_\mathrm{YSR}=E_o-E_e$. 

\begin{figure*}[t!]
         \centering
         \includegraphics[width=.75\textwidth]{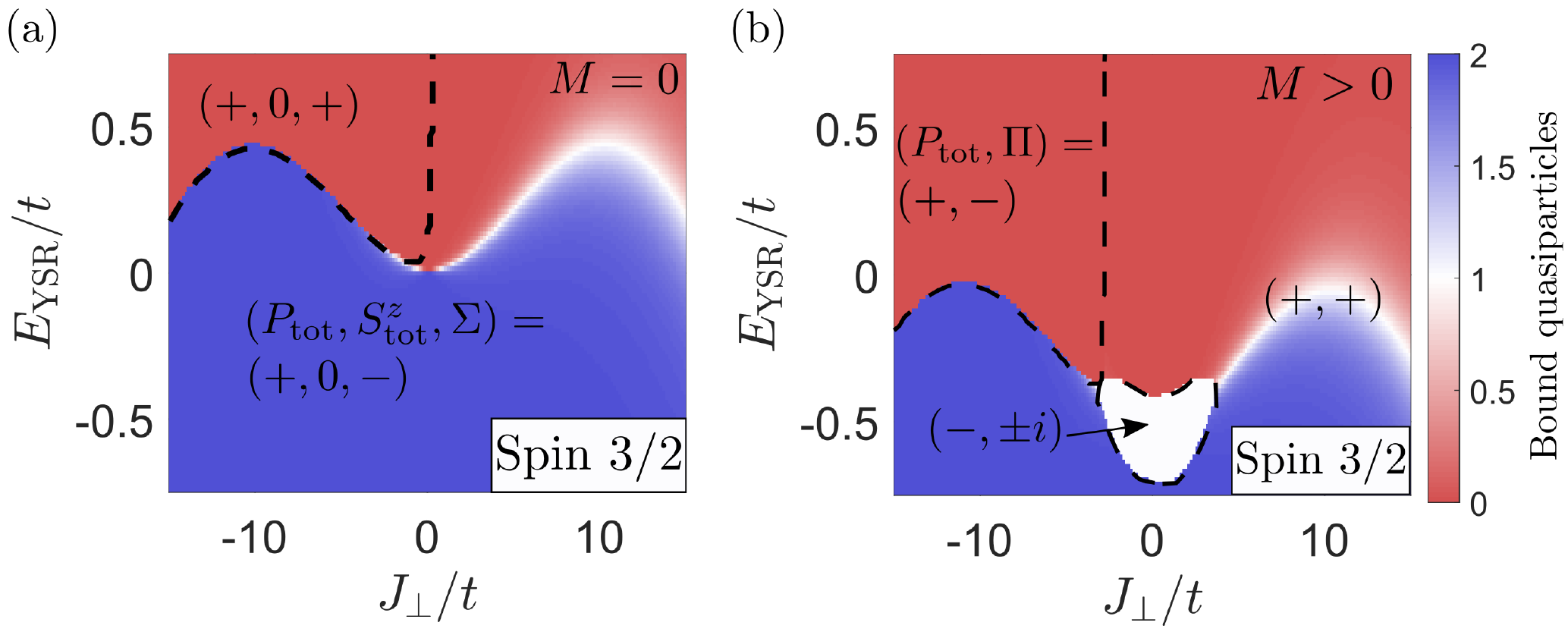}
     \caption{Phase diagrams for a spin-$\frac{3}{2}$ dimer with transverse ($xy$) couplings $K_\perp$ and $J_\perp$, easy-axis anisotropy $D$, and DM interaction $M$, see Sec.~\ref{subsec:anisotropic couplings}, in the intermediate regime $S \abs{D} \ll K_{\perp} \lesssim 4S^3 \abs{D}$. (a) $M=0$. Note that in contrast to Fig.\ \ref{fig_Yazdani}(b) transverse RKKY coupling stabilizes the doubly-screened dimer as the energy balance inverts for $K_{\perp} \lesssim 4S^4 \abs{D}$. (b) $M=3t$. In contrast to Fig.\ \ref{fig_Yazdani}(e) a singly-screened phase (white) arises. This requires  $K_{\perp} \lesssim 4S^3 \abs{D}$, see App. \ref{app:higher_spin_dimer_iso_dmi}.
     Parameters: $\Delta=200t$, $V=2\Delta$, 
     $D=-15t$. }
	\label{fig_Yazdani_Spin_32}
\end{figure*} 

\subsection{Phase diagrams at small $\abs{J}$}
\label{app:higher_spin_dimer_iso_phase}

There are subtleties for weak RKKY coupling $J$, which are briefly mentioned in the main text. In general, both the hybridization $\tilde t$ and the pairing $\tilde\Delta$ depend on $S_{\textrm{tot}}$. For instance, in the unscreened phases, $\tilde\Delta$ will vanish for sufficiently strong ferromagnetic RKKY coupling (maximal $S_\mathrm{tot}$), but is generally nonzero for antiferromagnetic coupling (minimal $S_\mathrm{tot}$). Close to $J = 0$, however, it is not a priori clear which $S_{\textrm{tot}}$ minimizes the total energy. $\tilde t$ favors a partially screened phase, $J$ favors ferromagnetic or antiferromagnetic coupling depending on its sign, and the singlet pairing $\tilde\Delta$ favors a singlet phase. The results of this competition are illustrated in Fig.\ \ref{fig:higher_spin_phase_diagrams_iso_app} for $S=2$. Note that these phase diagrams show a considerably smaller $J$ interval than the phase diagrams in the main text. 

The phase diagram depends sensitively on the relative magnitude of $\tilde t$ and $\tilde \Delta$ as controlled by the ratio $V/\Delta$. As we have seen in the main text, the available spin is typically maximized (minimized) for ferromagnetic (antiferromagnetic) RKKY coupling for $\Delta \sim V$ (so that $\tilde{t} \sim \tilde{\Delta}$) and the phase boundaries between ferromagnetic and antiferromagnetic phases occur at $J\simeq 0$. 

In contrast, for $V \gg \Delta$ and thus $\tilde t\gg \tilde\Delta$ [Fig.\ \ref{fig:higher_spin_phase_diagrams_iso_app}(a) and (b)], the singly screened phase extends into the region of antiferromagnetic coupling $J$ (up to $J \sim \tilde{t}$). In this region, there is 
a cascade of transitions from $S_{\textrm{tot}} = 2S - 1/2$ all the way down to $S_{\textrm{tot}} = 1/2$ in steps of one. (Some steps would only be apparent with higher resolution.) 

For $\Delta \gg V$ and thus $\tilde \Delta \gg \tilde t$ [Fig.\ \ref{fig:higher_spin_phase_diagrams_iso_app}(c) and (d)], the doubly screened phase $S_{\textrm{tot}} = 2S - 1$ as well as the antiferromagnetic phase gain energy via $\tilde{\Delta}$, reducing the extent of the singly screened phase. 

\subsection{Effective RKKY coupling}
\label{app:higher_spin_dimer_iso_RKKY}

Unlike for spin-$\frac{1}{2}$ adatoms, the RKKY coupling of higher spins is also nonzero for the (partially) screened states, in which the monomers now have a nonzero effective spin. To calculate the RKKY energy, we note that in the screened state, the matrix elements of the adatom spin $\mathbf{S}$ are proportional to the matrix elements of the effective adatom spin $\mathbf{S}_\mathrm{eff}$. As a consequence of the projection theorem, the constant of proportionality depends only on the magnitude of the effective spin $\mathbf{S}_\mathrm{eff}$ defined in Eq.\ (\ref{eq_seff}), so that we can replace
\begin{equation}
    \mathbf{S} \to c \mathbf{S}_\mathrm{eff}
\end{equation}
in the RKKY interaction,
\begin{equation}
    \mathbf{S}_1\cdot \hat J \cdot \mathbf{S}_2 \to
       c_1 c_2  \mathbf{S}_{\mathrm{eff},1}\cdot \hat J \cdot \mathbf{S}_{\mathrm{eff},2}.
\end{equation}
The constant $c_j$ is equal to unity if the monomer is in the unscreened state with $S_\mathrm{eff} =S$ and $c_j=1+\frac{1}{2S+1}$ if the monomer is in the screened state with $S_\mathrm{eff}=S-\frac{1}{2}$. In computing the RKKY coupling, we can then work with the effective adatom spin provided that we use the renormalized RKKY coupling $c_1 c_2 \hat J$.  

It remains to compute $c$ in the screened state. The
projection theorem gives
\begin{align}
\bra{S_{\textrm{eff}}^z} \mathbf{S} \ket{(S_{\textrm{eff}}^z)'}=&\ \frac{\bra{S_{\textrm{eff}}^z} \mathbf{S} \cdot \mathbf{S}_{\mathrm{eff}}\ket{S_{\textrm{eff}}^z}}{(S-\frac{1}{2})(S+\frac{1}{2})}  \bra{S_{\textrm{eff}}^z} \mathbf{S}_{\mathrm{eff}} \ket{(S_{\textrm{eff}}^z)'},
\end{align}
where $\ket{S_{\textrm{eff}}^z}$ denotes the projections of the effective spin $\mathbf{S}_\mathrm{eff}$ for $S_\mathrm{eff}=S-\frac{1}{2}$. Since $\mathbf{S} \cdot \mathbf{S}_{\mathrm{eff}}=(\mathbf{S}_{\mathrm{eff}}^2+\mathbf{S}^2- \mathbf{s}^2)/2$, the prefactor can be evaluated to give
\begin{align}
\bra{S^z_{\textrm{eff}}} \mathbf{S} \ket{(S_{\textrm{eff}}^z)'}=&\  \left(1+\frac{1}{2S+1}\right) \bra{S^z_{\textrm{eff}}} \mathbf{S}_{\mathrm{eff}} \ket{(S_{\textrm{eff}}^z)'}
\end{align}
as advertised above.

\subsection{Crossover between Ising and Heisenberg exchange}
\label{app:higher_spin_dimer_exchange_aniso}

As the ratio $K_z/K_\perp$ increases, the phase diagrams should become more and more classical. We probe this crossover in the limit of large $S$ by calculating how the phase boundaries and crossover lines in Fig.\ \ref{fig:higher_spin_phase_diagrams_iso} are modified when $K_z/K_\perp$ is different from unity. We focus on dominantly longitudinal exchange coupling, $K_z/K_\perp>1$, so that the screened monomers have maximal spin projection $S^z_\mathrm{eff} = \pm (S -\frac{1}{2})$. For simplicity, we assume $K_z-K_\perp \gg t,\abs{J}$ and work to leading order in the limit $S \gg 1$, which is expected to be most classical. 

First consider ferromagnetic RKKY coupling, for which we find the RKKY energies 
\begin{align}
    E_\mathrm{RKKY} \simeq J
    \begin{cases}
     S^2 & S^z_{\textrm{tot}} = \pm 2S,
     \\
     S\pqty{S - d_-^2 } & S^z_{\textrm{tot}} = \pm (2S - \frac{1}{2}),
     \\ 
     \pqty{S - d_-^2 }^2 & S^z_{\textrm{tot}} = \pm (2S - 1).
    \end{cases}
\end{align}
Here, $d_- = d_- (S - \frac{1}{2})$ is given by Eq.\ (\ref{dplusminus}) with $D=0$, which gives $d^2_- \simeq K_\perp^2/(2SK_z^2)$ for large $S$. By equating energies, we find that all three phase boundaries on the ferromagnetic side of the phase diagram follow
\begin{equation}
    E_{\textrm{YSR}} \simeq J S d_-^2 \simeq  \frac{J}{2} \frac{K_{\perp}^2}{K_z^2}.
\end{equation}
Thus, the phase boundary becomes classical only in the limit $K_\perp \ll K_z$. This is in contrast to spin$-\frac{1}{2}$ dimers, for which the classical behavior requires the even stronger condition  $K_{\perp} \ll t,J$. 
For antiferromagnetic RKKY coupling, the unscreened dimer gains RKKY energy $-JS(S+1)$, compared to $-J(S-d_-^2)^2$ for the doubly-screened dimer. This gives a crossover line 
\begin{equation}
E_{\textrm{YSR}} \simeq - J S\pqty{\frac{1}{2} + d_-^2} \simeq -JS/2
\end{equation}
in the limit of large $S$. Thus, for Heisenberg RKKY interaction, the slope of the crossover line is independent of the anisotropy of the exchange coupling. One finds classical behavior only when $J_z \gg J_{\perp}$ in addition to $K_z \gg K_{\perp}$. 

\subsection{DM interaction}
\label{app:higher_spin_dimer_iso_dmi}

We complement the semiclassical argument for the stabilization of the singly-screened phase due to DM interactions in Sec.\ \ref{subsec:anisotropic couplings} by a quantum-mechanical calculation. To this end, we estimate the energy gain due to the DM interaction in the various phases at small $J_\perp$.

First consider integer $S$, so that unscreened monomers have $S^z_\mathrm{eff} = \pm S$ and screened monomers $S^z_\mathrm{eff} = \pm \frac{1}{2}$. In the unscreened phase, the DM interaction $M(S^z_1 S^x_2 - S^x_1 S^z_2)$ lowers the energy only in quadratic order in $M$. In the doubly-screened phase, the DM interaction $M(S^z_1 S^x_2 - S^x_1 S^z_2)$ couples dimer states with $S^z_\mathrm{tot} = \pm 1$ (parallel monomer spins) to states with $S^z_\mathrm{tot} = 0$ (antiparallel monomer spins). The corresponding matrix elements involve
\begin{equation}
    \bra*{\frac{1}{2}, \frac{1}{2}}
     S^z_1 S^x_2 \ket*{\frac{1}{2},-\frac{1}{2}} =  \frac{1}{4}\sqrt{ S(S+1)} \sim S. 
\end{equation}
These results should be contrasted with the fact that in the partially-screened phase, the DM interaction couples monomer states with $S^z_\mathrm{tot} = \pm (S + \frac{1}{2})$ to states with $S^z_\mathrm{tot} = \pm (S - \frac{1}{2})$, with corresponding matrix elements 
\begin{equation}
    \bra*{S, \frac{1}{2}}
     S^z_1 S^x_2 \ket*{S,-\frac{1}{2}} =  \frac{1}{2} S \sqrt{ S(S+1)} \sim S^2. 
\end{equation}
Thus, the gain in DM energy scales as $MS^2$ in the singly-screened phase, compared to $MS$ for the doubly-screened phase and a yet smaller result in the unscreened phase. This explains the emergence of the odd-parity phase due to DM interactions in Figs.\ \ref{fig_Yazdani}(d) and (f) for integer spin. 

The situation is more subtle for half-integer $S$. Now, the screened monomers have $S^z_\mathrm{eff} = 0$. Thus, the DM interaction only couples to excited states involving monomer energies $\sim \abs{D}$ (for unscreened monomers) or $\sim K_{\perp}$ (for screened monomers). Correspondingly, the gain in DM energy approximately vanishes in the doubly-screened phase and equals $-M^2S^2/2\abs{D}$ in the unscreened phase. This should be compared to a gain in DM energy of $-2M^2S^5/K_\perp$ in the singly-screened phase. Thus, there is no singly-screened phase when the exchange coupling $K_\perp$ dominates over the other energy scales, as is the case in Fig.\ \ref{fig_Yazdani}(e). If, however, $4S^3 \abs{D} \gtrsim K_{\perp} \gg S \abs{D}$, a singly screened phase forms at small $J_{\perp}$, as shown in Fig. \ref{fig_Yazdani_Spin_32}(b).

\bibliographystyle{apsrev4-2}


\begin{thebibliography}{57}%
\makeatletter
\providecommand \@ifxundefined [1]{%
 \@ifx{#1\undefined}
}%
\providecommand \@ifnum [1]{%
 \ifnum #1\expandafter \@firstoftwo
 \else \expandafter \@secondoftwo
 \fi
}%
\providecommand \@ifx [1]{%
 \ifx #1\expandafter \@firstoftwo
 \else \expandafter \@secondoftwo
 \fi
}%
\providecommand \natexlab [1]{#1}%
\providecommand \enquote  [1]{``#1''}%
\providecommand \bibnamefont  [1]{#1}%
\providecommand \bibfnamefont [1]{#1}%
\providecommand \citenamefont [1]{#1}%
\providecommand \href@noop [0]{\@secondoftwo}%
\providecommand \href [0]{\begingroup \@sanitize@url \@href}%
\providecommand \@href[1]{\@@startlink{#1}\@@href}%
\providecommand \@@href[1]{\endgroup#1\@@endlink}%
\providecommand \@sanitize@url [0]{\catcode `\\12\catcode `\$12\catcode
  `\&12\catcode `\#12\catcode `\^12\catcode `\_12\catcode `\%12\relax}%
\providecommand \@@startlink[1]{}%
\providecommand \@@endlink[0]{}%
\providecommand \url  [0]{\begingroup\@sanitize@url \@url }%
\providecommand \@url [1]{\endgroup\@href {#1}{\urlprefix }}%
\providecommand \urlprefix  [0]{URL }%
\providecommand \Eprint [0]{\href }%
\providecommand \doibase [0]{https://doi.org/}%
\providecommand \selectlanguage [0]{\@gobble}%
\providecommand \bibinfo  [0]{\@secondoftwo}%
\providecommand \bibfield  [0]{\@secondoftwo}%
\providecommand \translation [1]{[#1]}%
\providecommand \BibitemOpen [0]{}%
\providecommand \bibitemStop [0]{}%
\providecommand \bibitemNoStop [0]{.\EOS\space}%
\providecommand \EOS [0]{\spacefactor3000\relax}%
\providecommand \BibitemShut  [1]{\csname bibitem#1\endcsname}%
\let\auto@bib@innerbib\@empty
\bibitem [{\citenamefont {Pawlak}\ \emph {et~al.}(2019)\citenamefont {Pawlak},
  \citenamefont {Hoffman}, \citenamefont {Klinovaja}, \citenamefont {Loss},\
  and\ \citenamefont {Meyer}}]{Pawlak2019}%
  \BibitemOpen
  \bibfield  {author} {\bibinfo {author} {\bibfnamefont {R.}~\bibnamefont
  {Pawlak}}, \bibinfo {author} {\bibfnamefont {S.}~\bibnamefont {Hoffman}},
  \bibinfo {author} {\bibfnamefont {J.}~\bibnamefont {Klinovaja}}, \bibinfo
  {author} {\bibfnamefont {D.}~\bibnamefont {Loss}},\ and\ \bibinfo {author}
  {\bibfnamefont {E.}~\bibnamefont {Meyer}},\ }\href
  {https://doi.org/10.1016/j.ppnp.2019.04.004} {\bibfield  {journal} {\bibinfo
  {journal} {Prog. Part. Nucl. Phys.}\ }\textbf {\bibinfo {volume} {107}},\
  \bibinfo {pages} {1} (\bibinfo {year} {2019})}\BibitemShut {NoStop}%
\bibitem [{\citenamefont {J\"ack}\ \emph {et~al.}(2021)\citenamefont {J\"ack},
  \citenamefont {Xie},\ and\ \citenamefont {Yazdani}}]{Jack2021}%
  \BibitemOpen
  \bibfield  {author} {\bibinfo {author} {\bibfnamefont {B.}~\bibnamefont
  {J\"ack}}, \bibinfo {author} {\bibfnamefont {Y.}~\bibnamefont {Xie}},\ and\
  \bibinfo {author} {\bibfnamefont {A.}~\bibnamefont {Yazdani}},\ }\href
  {https://www.nature.com/articles/s42254-021-00328-z} {\bibfield  {journal}
  {\bibinfo  {journal} {Nat. Rev. Phys.}\ }\textbf {\bibinfo {volume} {3}},\
  \bibinfo {pages} {541} (\bibinfo {year} {2021})}\BibitemShut {NoStop}%
\bibitem [{\citenamefont {Flensberg}\ \emph {et~al.}(2021)\citenamefont
  {Flensberg}, \citenamefont {von Oppen},\ and\ \citenamefont
  {Stern}}]{Flensberg2021}%
  \BibitemOpen
  \bibfield  {author} {\bibinfo {author} {\bibfnamefont {K.}~\bibnamefont
  {Flensberg}}, \bibinfo {author} {\bibfnamefont {F.}~\bibnamefont {von
  Oppen}},\ and\ \bibinfo {author} {\bibfnamefont {A.}~\bibnamefont {Stern}},\
  }\href {https://doi.org/10.1038/s41578-021-00336-6} {\bibfield  {journal}
  {\bibinfo  {journal} {Nat. Rev. Mater.}\ }\textbf {\bibinfo {volume} {6}},\
  \bibinfo {pages} {944} (\bibinfo {year} {2021})}\BibitemShut {NoStop}%
\bibitem [{\citenamefont {Steiner}\ \emph {et~al.}(2022)\citenamefont
  {Steiner}, \citenamefont {Mora}, \citenamefont {Franke},\ and\ \citenamefont
  {von Oppen}}]{Steiner2021}%
  \BibitemOpen
  \bibfield  {author} {\bibinfo {author} {\bibfnamefont {J.~F.}\ \bibnamefont
  {Steiner}}, \bibinfo {author} {\bibfnamefont {C.}~\bibnamefont {Mora}},
  \bibinfo {author} {\bibfnamefont {K.~J.}\ \bibnamefont {Franke}},\ and\
  \bibinfo {author} {\bibfnamefont {F.}~\bibnamefont {von Oppen}},\ }\href
  {https://doi.org/10.1103/PhysRevLett.128.036801} {\bibfield  {journal}
  {\bibinfo  {journal} {Phys. Rev. Lett.}\ }\textbf {\bibinfo {volume} {128}},\
  \bibinfo {pages} {036801} (\bibinfo {year} {2022})}\BibitemShut {NoStop}%
\bibitem [{\citenamefont {Yu}(1965)}]{Yu1965}%
  \BibitemOpen
  \bibfield  {author} {\bibinfo {author} {\bibfnamefont {L.}~\bibnamefont
  {Yu}},\ }\href {https://doi.org/10.7498/aps.21.75} {\bibfield  {journal}
  {\bibinfo  {journal} {Acta Phys. Sin.}\ }\textbf {\bibinfo {volume} {21}},\
  \bibinfo {pages} {75} (\bibinfo {year} {1965})}\BibitemShut {NoStop}%
\bibitem [{\citenamefont {Shiba}(1968)}]{Shiba1968}%
  \BibitemOpen
  \bibfield  {author} {\bibinfo {author} {\bibfnamefont {H.}~\bibnamefont
  {Shiba}},\ }\href {https://doi.org/10.1143/PTP.40.435} {\bibfield  {journal}
  {\bibinfo  {journal} {Prog. Theor. Phys.}\ }\textbf {\bibinfo {volume}
  {40}},\ \bibinfo {pages} {435} (\bibinfo {year} {1968})}\BibitemShut
  {NoStop}%
\bibitem [{\citenamefont {Rusinov}(1969)}]{Rusinov1969}%
  \BibitemOpen
  \bibfield  {author} {\bibinfo {author} {\bibfnamefont {A.~I.}\ \bibnamefont
  {Rusinov}},\ }\href {http://jetpletters.ru/ps/1658/article_25295.shtml}
  {\bibfield  {journal} {\bibinfo  {journal} {JETP Lett.}\ }\textbf {\bibinfo
  {volume} {9}},\ \bibinfo {pages} {85} (\bibinfo {year} {1969})}\BibitemShut
  {NoStop}%
\bibitem [{\citenamefont {Yazdani}\ \emph {et~al.}(1997)\citenamefont
  {Yazdani}, \citenamefont {Jones}, \citenamefont {Lutz}, \citenamefont
  {Crommie},\ and\ \citenamefont {Eigler}}]{Yazdani1997}%
  \BibitemOpen
  \bibfield  {author} {\bibinfo {author} {\bibfnamefont {A.}~\bibnamefont
  {Yazdani}}, \bibinfo {author} {\bibfnamefont {B.~A.}\ \bibnamefont {Jones}},
  \bibinfo {author} {\bibfnamefont {C.~P.}\ \bibnamefont {Lutz}}, \bibinfo
  {author} {\bibfnamefont {M.~F.}\ \bibnamefont {Crommie}},\ and\ \bibinfo
  {author} {\bibfnamefont {D.~M.}\ \bibnamefont {Eigler}},\ }\href
  {https://doi.org/10.1126/science.275.5307.1767} {\bibfield  {journal}
  {\bibinfo  {journal} {Science}\ }\textbf {\bibinfo {volume} {275}},\ \bibinfo
  {pages} {1767} (\bibinfo {year} {1997})}\BibitemShut {NoStop}%
\bibitem [{\citenamefont {Ji}\ \emph {et~al.}(2008)\citenamefont {Ji},
  \citenamefont {Zhang}, \citenamefont {Fu}, \citenamefont {Chen},
  \citenamefont {Ma}, \citenamefont {Li}, \citenamefont {Duan}, \citenamefont
  {Jia},\ and\ \citenamefont {Xue}}]{Ji2008}%
  \BibitemOpen
  \bibfield  {author} {\bibinfo {author} {\bibfnamefont {S.-H.}\ \bibnamefont
  {Ji}}, \bibinfo {author} {\bibfnamefont {T.}~\bibnamefont {Zhang}}, \bibinfo
  {author} {\bibfnamefont {Y.-S.}\ \bibnamefont {Fu}}, \bibinfo {author}
  {\bibfnamefont {X.}~\bibnamefont {Chen}}, \bibinfo {author} {\bibfnamefont
  {X.-C.}\ \bibnamefont {Ma}}, \bibinfo {author} {\bibfnamefont
  {J.}~\bibnamefont {Li}}, \bibinfo {author} {\bibfnamefont {W.-H.}\
  \bibnamefont {Duan}}, \bibinfo {author} {\bibfnamefont {J.-F.}\ \bibnamefont
  {Jia}},\ and\ \bibinfo {author} {\bibfnamefont {Q.-K.}\ \bibnamefont {Xue}},\
  }\href {https://doi.org/10.1103/PhysRevLett.100.226801} {\bibfield  {journal}
  {\bibinfo  {journal} {Phys. Rev. Lett.}\ }\textbf {\bibinfo {volume} {100}},\
  \bibinfo {pages} {226801} (\bibinfo {year} {2008})}\BibitemShut {NoStop}%
\bibitem [{\citenamefont {Franke}\ \emph {et~al.}(2011)\citenamefont {Franke},
  \citenamefont {Schulze},\ and\ \citenamefont {Pascual}}]{Franke2011}%
  \BibitemOpen
  \bibfield  {author} {\bibinfo {author} {\bibfnamefont {K.~J.}\ \bibnamefont
  {Franke}}, \bibinfo {author} {\bibfnamefont {G.}~\bibnamefont {Schulze}},\
  and\ \bibinfo {author} {\bibfnamefont {J.~I.}\ \bibnamefont {Pascual}},\
  }\href {https://doi.org/10.1126/science.1202204} {\bibfield  {journal}
  {\bibinfo  {journal} {Science}\ }\textbf {\bibinfo {volume} {332}},\ \bibinfo
  {pages} {940} (\bibinfo {year} {2011})}\BibitemShut {NoStop}%
\bibitem [{\citenamefont {Balatsky}\ \emph {et~al.}(2006)\citenamefont
  {Balatsky}, \citenamefont {Vekhter},\ and\ \citenamefont
  {Zhu}}]{Balatsky2006}%
  \BibitemOpen
  \bibfield  {author} {\bibinfo {author} {\bibfnamefont {A.~V.}\ \bibnamefont
  {Balatsky}}, \bibinfo {author} {\bibfnamefont {I.}~\bibnamefont {Vekhter}},\
  and\ \bibinfo {author} {\bibfnamefont {J.-X.}\ \bibnamefont {Zhu}},\ }\href
  {https://doi.org/10.1103/RevModPhys.78.373} {\bibfield  {journal} {\bibinfo
  {journal} {Rev. Mod. Phys.}\ }\textbf {\bibinfo {volume} {78}},\ \bibinfo
  {pages} {373} (\bibinfo {year} {2006})}\BibitemShut {NoStop}%
\bibitem [{\citenamefont {Heinrich}\ \emph {et~al.}(2018)\citenamefont
  {Heinrich}, \citenamefont {Pascual},\ and\ \citenamefont
  {Franke}}]{Heinrich2018}%
  \BibitemOpen
  \bibfield  {author} {\bibinfo {author} {\bibfnamefont {B.~W.}\ \bibnamefont
  {Heinrich}}, \bibinfo {author} {\bibfnamefont {J.~I.}\ \bibnamefont
  {Pascual}},\ and\ \bibinfo {author} {\bibfnamefont {K.~J.}\ \bibnamefont
  {Franke}},\ }\href {https://doi.org/10.1016/j.progsurf.2018.01.001}
  {\bibfield  {journal} {\bibinfo  {journal} {Prog. Surf. Sci.}\ }\textbf
  {\bibinfo {volume} {93}},\ \bibinfo {pages} {1} (\bibinfo {year}
  {2018})}\BibitemShut {NoStop}%
\bibitem [{\citenamefont {Ruby}\ \emph {et~al.}(2018)\citenamefont {Ruby},
  \citenamefont {Heinrich}, \citenamefont {Peng}, \citenamefont {von Oppen},\
  and\ \citenamefont {Franke}}]{Ruby2018}%
  \BibitemOpen
  \bibfield  {author} {\bibinfo {author} {\bibfnamefont {M.}~\bibnamefont
  {Ruby}}, \bibinfo {author} {\bibfnamefont {B.~W.}\ \bibnamefont {Heinrich}},
  \bibinfo {author} {\bibfnamefont {Y.}~\bibnamefont {Peng}}, \bibinfo {author}
  {\bibfnamefont {F.}~\bibnamefont {von Oppen}},\ and\ \bibinfo {author}
  {\bibfnamefont {K.~J.}\ \bibnamefont {Franke}},\ }\href
  {https://doi.org/10.1103/PhysRevLett.120.156803} {\bibfield  {journal}
  {\bibinfo  {journal} {Phys. Rev. Lett.}\ }\textbf {\bibinfo {volume} {120}},\
  \bibinfo {pages} {156803} (\bibinfo {year} {2018})}\BibitemShut {NoStop}%
\bibitem [{\citenamefont {Choi}\ \emph {et~al.}(2018)\citenamefont {Choi},
  \citenamefont {Fern\'andez}, \citenamefont {Herrera}, \citenamefont
  {Rubio-Verd{\'{u}}}, \citenamefont {Ugeda}, \citenamefont {Guillam\'on},
  \citenamefont {Suderow}, \citenamefont {Pascual},\ and\ \citenamefont
  {Lorente}}]{Choi2018}%
  \BibitemOpen
  \bibfield  {author} {\bibinfo {author} {\bibfnamefont {D.-J.}\ \bibnamefont
  {Choi}}, \bibinfo {author} {\bibfnamefont {C.~G.}\ \bibnamefont
  {Fern\'andez}}, \bibinfo {author} {\bibfnamefont {E.}~\bibnamefont
  {Herrera}}, \bibinfo {author} {\bibfnamefont {C.}~\bibnamefont
  {Rubio-Verd{\'{u}}}}, \bibinfo {author} {\bibfnamefont {M.~M.}\ \bibnamefont
  {Ugeda}}, \bibinfo {author} {\bibfnamefont {I.}~\bibnamefont {Guillam\'on}},
  \bibinfo {author} {\bibfnamefont {H.}~\bibnamefont {Suderow}}, \bibinfo
  {author} {\bibfnamefont {J.~I.}\ \bibnamefont {Pascual}},\ and\ \bibinfo
  {author} {\bibfnamefont {N.}~\bibnamefont {Lorente}},\ }\href
  {https://doi.org/10.1103/PhysRevLett.120.167001} {\bibfield  {journal}
  {\bibinfo  {journal} {Phys. Rev. Lett.}\ }\textbf {\bibinfo {volume} {120}},\
  \bibinfo {pages} {167001} (\bibinfo {year} {2018})}\BibitemShut {NoStop}%
\bibitem [{\citenamefont {Kezilebieke}\ \emph {et~al.}(2018)\citenamefont
  {Kezilebieke}, \citenamefont {Dvorak}, \citenamefont {Ojanen},\ and\
  \citenamefont {Liljeroth}}]{Kezilebieke2018}%
  \BibitemOpen
  \bibfield  {author} {\bibinfo {author} {\bibfnamefont {S.}~\bibnamefont
  {Kezilebieke}}, \bibinfo {author} {\bibfnamefont {M.}~\bibnamefont {Dvorak}},
  \bibinfo {author} {\bibfnamefont {T.}~\bibnamefont {Ojanen}},\ and\ \bibinfo
  {author} {\bibfnamefont {P.}~\bibnamefont {Liljeroth}},\ }\href
  {https://doi.org/10.1021/acs.nanolett.7b05050} {\bibfield  {journal}
  {\bibinfo  {journal} {Nano Lett.}\ }\textbf {\bibinfo {volume} {18}},\
  \bibinfo {pages} {2311} (\bibinfo {year} {2018})}\BibitemShut {NoStop}%
\bibitem [{\citenamefont {Beck}\ \emph {et~al.}(2021)\citenamefont {Beck},
  \citenamefont {Schneider}, \citenamefont {Rózsa}, \citenamefont {Palotás},
  \citenamefont {Lászlóffy}, \citenamefont {Szunyogh}, \citenamefont
  {Wiebe},\ and\ \citenamefont {Wiesendanger}}]{Beck2020}%
  \BibitemOpen
  \bibfield  {author} {\bibinfo {author} {\bibfnamefont {P.}~\bibnamefont
  {Beck}}, \bibinfo {author} {\bibfnamefont {L.}~\bibnamefont {Schneider}},
  \bibinfo {author} {\bibfnamefont {L.}~\bibnamefont {Rózsa}}, \bibinfo
  {author} {\bibfnamefont {K.}~\bibnamefont {Palotás}}, \bibinfo {author}
  {\bibfnamefont {A.}~\bibnamefont {Lászlóffy}}, \bibinfo {author}
  {\bibfnamefont {L.}~\bibnamefont {Szunyogh}}, \bibinfo {author}
  {\bibfnamefont {J.}~\bibnamefont {Wiebe}},\ and\ \bibinfo {author}
  {\bibfnamefont {R.}~\bibnamefont {Wiesendanger}},\ }\href
  {https://doi.org/10.1038/s41467-021-22261-6} {\bibfield  {journal} {\bibinfo
  {journal} {Nat. Commun.}\ }\textbf {\bibinfo {volume} {12}},\ \bibinfo
  {pages} {2040} (\bibinfo {year} {2021})}\BibitemShut {NoStop}%
\bibitem [{\citenamefont {Ding}\ \emph {et~al.}(2021)\citenamefont {Ding},
  \citenamefont {Hu}, \citenamefont {Randeria}, \citenamefont {Hoffman},
  \citenamefont {Deb}, \citenamefont {Klinovaja}, \citenamefont {Loss},\ and\
  \citenamefont {Yazdani}}]{Ding2021}%
  \BibitemOpen
  \bibfield  {author} {\bibinfo {author} {\bibfnamefont {H.}~\bibnamefont
  {Ding}}, \bibinfo {author} {\bibfnamefont {Y.}~\bibnamefont {Hu}}, \bibinfo
  {author} {\bibfnamefont {M.~T.}\ \bibnamefont {Randeria}}, \bibinfo {author}
  {\bibfnamefont {S.}~\bibnamefont {Hoffman}}, \bibinfo {author} {\bibfnamefont
  {O.}~\bibnamefont {Deb}}, \bibinfo {author} {\bibfnamefont {J.}~\bibnamefont
  {Klinovaja}}, \bibinfo {author} {\bibfnamefont {D.}~\bibnamefont {Loss}},\
  and\ \bibinfo {author} {\bibfnamefont {A.}~\bibnamefont {Yazdani}},\ }\href
  {https://doi.org/10.1073/pnas.2024837118} {\bibfield  {journal} {\bibinfo
  {journal} {Proc. Nat. Acad. Sci.}\ }\textbf {\bibinfo {volume} {118}},\
  \bibinfo {pages} {e2024837118} (\bibinfo {year} {2021})}\BibitemShut
  {NoStop}%
\bibitem [{\citenamefont {Kamlapure}\ \emph {et~al.}(2018)\citenamefont
  {Kamlapure}, \citenamefont {Cornils}, \citenamefont {Wiebe},\ and\
  \citenamefont {Wiesendanger}}]{Kamlapure2018}%
  \BibitemOpen
  \bibfield  {author} {\bibinfo {author} {\bibfnamefont {A.}~\bibnamefont
  {Kamlapure}}, \bibinfo {author} {\bibfnamefont {L.}~\bibnamefont {Cornils}},
  \bibinfo {author} {\bibfnamefont {J.}~\bibnamefont {Wiebe}},\ and\ \bibinfo
  {author} {\bibfnamefont {R.}~\bibnamefont {Wiesendanger}},\ }\href
  {https://doi.org/10.1038/s41467-018-05701-8} {\bibfield  {journal} {\bibinfo
  {journal} {Nat. Commun.}\ }\textbf {\bibinfo {volume} {9}},\ \bibinfo {pages}
  {1} (\bibinfo {year} {2018})}\BibitemShut {NoStop}%
\bibitem [{\citenamefont {Küster}\ \emph {et~al.}(2021)\citenamefont
  {Küster}, \citenamefont {Brinker}, \citenamefont {Lounis}, \citenamefont
  {Parkin},\ and\ \citenamefont {Sessi}}]{Kuester2021}%
  \BibitemOpen
  \bibfield  {author} {\bibinfo {author} {\bibfnamefont {F.}~\bibnamefont
  {Küster}}, \bibinfo {author} {\bibfnamefont {S.}~\bibnamefont {Brinker}},
  \bibinfo {author} {\bibfnamefont {S.}~\bibnamefont {Lounis}}, \bibinfo
  {author} {\bibfnamefont {S.~S.~P.}\ \bibnamefont {Parkin}},\ and\ \bibinfo
  {author} {\bibfnamefont {P.}~\bibnamefont {Sessi}},\ }\href
  {https://www.nature.com/articles/s41467-021-26802-x} {\bibfield  {journal}
  {\bibinfo  {journal} {Nat. Commun.}\ }\textbf {\bibinfo {volume} {12}},\
  \bibinfo {pages} {6722} (\bibinfo {year} {2021})}\BibitemShut {NoStop}%
\bibitem [{\citenamefont {Liebhaber}\ \emph {et~al.}(2021)\citenamefont
  {Liebhaber}, \citenamefont {Rütten}, \citenamefont {Reecht}, \citenamefont
  {Steiner}, \citenamefont {Rohlf}, \citenamefont {Rossnagel}, \citenamefont
  {von Oppen},\ and\ \citenamefont {Franke}}]{Liebhaber2021}%
  \BibitemOpen
  \bibfield  {author} {\bibinfo {author} {\bibfnamefont {E.}~\bibnamefont
  {Liebhaber}}, \bibinfo {author} {\bibfnamefont {L.~M.}\ \bibnamefont
  {Rütten}}, \bibinfo {author} {\bibfnamefont {G.}~\bibnamefont {Reecht}},
  \bibinfo {author} {\bibfnamefont {J.~F.}\ \bibnamefont {Steiner}}, \bibinfo
  {author} {\bibfnamefont {S.}~\bibnamefont {Rohlf}}, \bibinfo {author}
  {\bibfnamefont {K.}~\bibnamefont {Rossnagel}}, \bibinfo {author}
  {\bibfnamefont {F.}~\bibnamefont {von Oppen}},\ and\ \bibinfo {author}
  {\bibfnamefont {K.~J.}\ \bibnamefont {Franke}},\ }\href@noop {} {\bibinfo
  {title} {Quantum spins and hybridization in artificially-constructed chains
  of magnetic adatoms on a superconductor}} (\bibinfo {year} {2021}),\ \Eprint
  {https://arxiv.org/abs/2107.06361} {arXiv:2107.06361} \BibitemShut {NoStop}%
\bibitem [{\citenamefont {Huang}\ \emph {et~al.}(2020)\citenamefont {Huang},
  \citenamefont {Padurariu}, \citenamefont {Senkpiel}, \citenamefont {Drost},
  \citenamefont {Yeyati}, \citenamefont {Cuevas}, \citenamefont {Kubala},
  \citenamefont {Ankerhold}, \citenamefont {Kern},\ and\ \citenamefont
  {Ast}}]{Huang2020}%
  \BibitemOpen
  \bibfield  {author} {\bibinfo {author} {\bibfnamefont {H.}~\bibnamefont
  {Huang}}, \bibinfo {author} {\bibfnamefont {C.}~\bibnamefont {Padurariu}},
  \bibinfo {author} {\bibfnamefont {J.}~\bibnamefont {Senkpiel}}, \bibinfo
  {author} {\bibfnamefont {R.}~\bibnamefont {Drost}}, \bibinfo {author}
  {\bibfnamefont {A.~L.}\ \bibnamefont {Yeyati}}, \bibinfo {author}
  {\bibfnamefont {J.~C.}\ \bibnamefont {Cuevas}}, \bibinfo {author}
  {\bibfnamefont {B.}~\bibnamefont {Kubala}}, \bibinfo {author} {\bibfnamefont
  {J.}~\bibnamefont {Ankerhold}}, \bibinfo {author} {\bibfnamefont
  {K.}~\bibnamefont {Kern}},\ and\ \bibinfo {author} {\bibfnamefont {C.~R.}\
  \bibnamefont {Ast}},\ }\href {https://doi.org/10.1038/s41567-020-0971-0}
  {\bibfield  {journal} {\bibinfo  {journal} {Nat. Phys.}\ }\textbf {\bibinfo
  {volume} {16}},\ \bibinfo {pages} {1227} (\bibinfo {year}
  {2020})}\BibitemShut {NoStop}%
\bibitem [{\citenamefont {Ruderman}\ and\ \citenamefont
  {Kittel}(1954)}]{Ruderman1954}%
  \BibitemOpen
  \bibfield  {author} {\bibinfo {author} {\bibfnamefont {M.~A.}\ \bibnamefont
  {Ruderman}}\ and\ \bibinfo {author} {\bibfnamefont {C.}~\bibnamefont
  {Kittel}},\ }\href {https://doi.org/10.1103/PhysRev.96.99} {\bibfield
  {journal} {\bibinfo  {journal} {Phys. Rev.}\ }\textbf {\bibinfo {volume}
  {96}},\ \bibinfo {pages} {99} (\bibinfo {year} {1954})}\BibitemShut {NoStop}%
\bibitem [{\citenamefont {{Kasuya}}(1956)}]{Kasuya1956}%
  \BibitemOpen
  \bibfield  {author} {\bibinfo {author} {\bibfnamefont {T.}~\bibnamefont
  {{Kasuya}}},\ }\href@noop {} {\bibfield  {journal} {\bibinfo  {journal}
  {{Prog. Theor. Phys.}}\ }\textbf {\bibinfo {volume} {16}},\ \bibinfo {pages}
  {45} (\bibinfo {year} {1956})}\BibitemShut {NoStop}%
\bibitem [{\citenamefont {Yosida}(1957)}]{Yosida1957}%
  \BibitemOpen
  \bibfield  {author} {\bibinfo {author} {\bibfnamefont {K.}~\bibnamefont
  {Yosida}},\ }\href {https://doi.org/10.1103/PhysRev.106.893} {\bibfield
  {journal} {\bibinfo  {journal} {Phys. Rev.}\ }\textbf {\bibinfo {volume}
  {106}},\ \bibinfo {pages} {893} (\bibinfo {year} {1957})}\BibitemShut
  {NoStop}%
\bibitem [{\citenamefont {Anderson}\ and\ \citenamefont
  {Suhl}(1959)}]{Anderson1959}%
  \BibitemOpen
  \bibfield  {author} {\bibinfo {author} {\bibfnamefont {P.~W.}\ \bibnamefont
  {Anderson}}\ and\ \bibinfo {author} {\bibfnamefont {H.}~\bibnamefont
  {Suhl}},\ }\href {https://doi.org/10.1103/PhysRev.116.898} {\bibfield
  {journal} {\bibinfo  {journal} {Phys. Rev.}\ }\textbf {\bibinfo {volume}
  {116}},\ \bibinfo {pages} {898} (\bibinfo {year} {1959})}\BibitemShut
  {NoStop}%
\bibitem [{\citenamefont {Dzyaloshinsky}(1958)}]{Dzyaloshinsky1958}%
  \BibitemOpen
  \bibfield  {author} {\bibinfo {author} {\bibfnamefont {I.}~\bibnamefont
  {Dzyaloshinsky}},\ }\href
  {https://doi.org/https://doi.org/10.1016/0022-3697(58)90076-3} {\bibfield
  {journal} {\bibinfo  {journal} {J. Phys. Chem. Solids}\ }\textbf {\bibinfo
  {volume} {4}},\ \bibinfo {pages} {241} (\bibinfo {year} {1958})}\BibitemShut
  {NoStop}%
\bibitem [{\citenamefont {Moriya}(1960)}]{Moriya1960}%
  \BibitemOpen
  \bibfield  {author} {\bibinfo {author} {\bibfnamefont {T.}~\bibnamefont
  {Moriya}},\ }\href {https://doi.org/10.1103/PhysRev.120.91} {\bibfield
  {journal} {\bibinfo  {journal} {Phys. Rev.}\ }\textbf {\bibinfo {volume}
  {120}},\ \bibinfo {pages} {91} (\bibinfo {year} {1960})}\BibitemShut
  {NoStop}%
\bibitem [{\citenamefont {Pientka}\ \emph {et~al.}(2013)\citenamefont
  {Pientka}, \citenamefont {Glazman},\ and\ \citenamefont {von
  Oppen}}]{Pientka2013}%
  \BibitemOpen
  \bibfield  {author} {\bibinfo {author} {\bibfnamefont {F.}~\bibnamefont
  {Pientka}}, \bibinfo {author} {\bibfnamefont {L.~I.}\ \bibnamefont
  {Glazman}},\ and\ \bibinfo {author} {\bibfnamefont {F.}~\bibnamefont {von
  Oppen}},\ }\href {https://doi.org/10.1103/PhysRevB.88.155420} {\bibfield
  {journal} {\bibinfo  {journal} {Phys. Rev. B}\ }\textbf {\bibinfo {volume}
  {88}},\ \bibinfo {pages} {155420} (\bibinfo {year} {2013})}\BibitemShut
  {NoStop}%
\bibitem [{\citenamefont {P\"oyh\"onen}\ \emph {et~al.}(2014)\citenamefont
  {P\"oyh\"onen}, \citenamefont {Weststr\"om}, \citenamefont {R\"ontynen},\
  and\ \citenamefont {Ojanen}}]{Poyhonen2014}%
  \BibitemOpen
  \bibfield  {author} {\bibinfo {author} {\bibfnamefont {K.}~\bibnamefont
  {P\"oyh\"onen}}, \bibinfo {author} {\bibfnamefont {A.}~\bibnamefont
  {Weststr\"om}}, \bibinfo {author} {\bibfnamefont {J.}~\bibnamefont
  {R\"ontynen}},\ and\ \bibinfo {author} {\bibfnamefont {T.}~\bibnamefont
  {Ojanen}},\ }\href {https://doi.org/10.1103/PhysRevB.89.115109} {\bibfield
  {journal} {\bibinfo  {journal} {Phys. Rev. B}\ }\textbf {\bibinfo {volume}
  {89}},\ \bibinfo {pages} {115109} (\bibinfo {year} {2014})}\BibitemShut
  {NoStop}%
\bibitem [{\citenamefont {Meng}\ \emph {et~al.}(2015)\citenamefont {Meng},
  \citenamefont {Klinovaja}, \citenamefont {Hoffman}, \citenamefont {Simon},\
  and\ \citenamefont {Loss}}]{Meng2015}%
  \BibitemOpen
  \bibfield  {author} {\bibinfo {author} {\bibfnamefont {T.}~\bibnamefont
  {Meng}}, \bibinfo {author} {\bibfnamefont {J.}~\bibnamefont {Klinovaja}},
  \bibinfo {author} {\bibfnamefont {S.}~\bibnamefont {Hoffman}}, \bibinfo
  {author} {\bibfnamefont {P.}~\bibnamefont {Simon}},\ and\ \bibinfo {author}
  {\bibfnamefont {D.}~\bibnamefont {Loss}},\ }\href
  {https://doi.org/10.1103/PhysRevB.92.064503} {\bibfield  {journal} {\bibinfo
  {journal} {Phys. Rev. B}\ }\textbf {\bibinfo {volume} {92}},\ \bibinfo
  {pages} {064503} (\bibinfo {year} {2015})}\BibitemShut {NoStop}%
\bibitem [{\citenamefont {Hoffman}\ \emph {et~al.}(2015)\citenamefont
  {Hoffman}, \citenamefont {Klinovaja}, \citenamefont {Meng},\ and\
  \citenamefont {Loss}}]{Hoffman2015}%
  \BibitemOpen
  \bibfield  {author} {\bibinfo {author} {\bibfnamefont {S.}~\bibnamefont
  {Hoffman}}, \bibinfo {author} {\bibfnamefont {J.}~\bibnamefont {Klinovaja}},
  \bibinfo {author} {\bibfnamefont {T.}~\bibnamefont {Meng}},\ and\ \bibinfo
  {author} {\bibfnamefont {D.}~\bibnamefont {Loss}},\ }\href
  {https://doi.org/10.1103/PhysRevB.92.125422} {\bibfield  {journal} {\bibinfo
  {journal} {Phys. Rev. B}\ }\textbf {\bibinfo {volume} {92}},\ \bibinfo
  {pages} {125422} (\bibinfo {year} {2015})}\BibitemShut {NoStop}%
\bibitem [{\citenamefont {Brydon}\ \emph {et~al.}(2015)\citenamefont {Brydon},
  \citenamefont {Das~Sarma}, \citenamefont {Hui},\ and\ \citenamefont
  {Sau}}]{Brydon2015}%
  \BibitemOpen
  \bibfield  {author} {\bibinfo {author} {\bibfnamefont {P.~M.~R.}\
  \bibnamefont {Brydon}}, \bibinfo {author} {\bibfnamefont {S.}~\bibnamefont
  {Das~Sarma}}, \bibinfo {author} {\bibfnamefont {H.-Y.}\ \bibnamefont {Hui}},\
  and\ \bibinfo {author} {\bibfnamefont {J.~D.}\ \bibnamefont {Sau}},\ }\href
  {https://doi.org/10.1103/PhysRevB.91.064505} {\bibfield  {journal} {\bibinfo
  {journal} {Phys. Rev. B}\ }\textbf {\bibinfo {volume} {91}},\ \bibinfo
  {pages} {064505} (\bibinfo {year} {2015})}\BibitemShut {NoStop}%
\bibitem [{\citenamefont {K\"orber}\ \emph {et~al.}(2018)\citenamefont
  {K\"orber}, \citenamefont {Trauzettel},\ and\ \citenamefont
  {Kashuba}}]{Korber2018}%
  \BibitemOpen
  \bibfield  {author} {\bibinfo {author} {\bibfnamefont {S.}~\bibnamefont
  {K\"orber}}, \bibinfo {author} {\bibfnamefont {B.}~\bibnamefont
  {Trauzettel}},\ and\ \bibinfo {author} {\bibfnamefont {O.}~\bibnamefont
  {Kashuba}},\ }\href {https://doi.org/10.1103/PhysRevB.97.184503} {\bibfield
  {journal} {\bibinfo  {journal} {Phys. Rev. B}\ }\textbf {\bibinfo {volume}
  {97}},\ \bibinfo {pages} {184503} (\bibinfo {year} {2018})}\BibitemShut
  {NoStop}%
\bibitem [{\citenamefont {Madhavan}\ \emph {et~al.}(1998)\citenamefont
  {Madhavan}, \citenamefont {Chen}, \citenamefont {Jamneala}, \citenamefont
  {Crommie},\ and\ \citenamefont {Wingreen}}]{Madhavan1998}%
  \BibitemOpen
  \bibfield  {author} {\bibinfo {author} {\bibfnamefont {V.}~\bibnamefont
  {Madhavan}}, \bibinfo {author} {\bibfnamefont {W.}~\bibnamefont {Chen}},
  \bibinfo {author} {\bibfnamefont {T.}~\bibnamefont {Jamneala}}, \bibinfo
  {author} {\bibfnamefont {M.~F.}\ \bibnamefont {Crommie}},\ and\ \bibinfo
  {author} {\bibfnamefont {N.~S.}\ \bibnamefont {Wingreen}},\ }\href
  {https://doi.org/10.1126/science.280.5363.567} {\bibfield  {journal}
  {\bibinfo  {journal} {Science}\ }\textbf {\bibinfo {volume} {280}},\ \bibinfo
  {pages} {567} (\bibinfo {year} {1998})}\BibitemShut {NoStop}%
\bibitem [{\citenamefont {Li}\ \emph {et~al.}(1998)\citenamefont {Li},
  \citenamefont {Schneider}, \citenamefont {Berndt},\ and\ \citenamefont
  {Delley}}]{Li1998}%
  \BibitemOpen
  \bibfield  {author} {\bibinfo {author} {\bibfnamefont {J.}~\bibnamefont
  {Li}}, \bibinfo {author} {\bibfnamefont {W.-D.}\ \bibnamefont {Schneider}},
  \bibinfo {author} {\bibfnamefont {R.}~\bibnamefont {Berndt}},\ and\ \bibinfo
  {author} {\bibfnamefont {B.}~\bibnamefont {Delley}},\ }\href
  {https://doi.org/10.1103/PhysRevLett.80.2893} {\bibfield  {journal} {\bibinfo
   {journal} {Phys. Rev. Lett.}\ }\textbf {\bibinfo {volume} {80}},\ \bibinfo
  {pages} {2893} (\bibinfo {year} {1998})}\BibitemShut {NoStop}%
\bibitem [{\citenamefont {Hatter}\ \emph {et~al.}(2017)\citenamefont {Hatter},
  \citenamefont {Heinrich}, \citenamefont {Rolf},\ and\ \citenamefont
  {Franke}}]{Hatter2017}%
  \BibitemOpen
  \bibfield  {author} {\bibinfo {author} {\bibfnamefont {N.}~\bibnamefont
  {Hatter}}, \bibinfo {author} {\bibfnamefont {B.~W.}\ \bibnamefont
  {Heinrich}}, \bibinfo {author} {\bibfnamefont {D.}~\bibnamefont {Rolf}},\
  and\ \bibinfo {author} {\bibfnamefont {K.~J.}\ \bibnamefont {Franke}},\
  }\href {https://doi.org/10.1038/s41467-017-02277-7} {\bibfield  {journal}
  {\bibinfo  {journal} {Nat. Commun.}\ }\textbf {\bibinfo {volume} {8}},\
  \bibinfo {pages} {2016} (\bibinfo {year} {2017})}\BibitemShut {NoStop}%
\bibitem [{\citenamefont {Farinacci}\ \emph {et~al.}(2020)\citenamefont
  {Farinacci}, \citenamefont {Ahmadi}, \citenamefont {Ruby}, \citenamefont
  {Reecht}, \citenamefont {Heinrich}, \citenamefont {Czekelius}, \citenamefont
  {von Oppen},\ and\ \citenamefont {Franke}}]{Farinacci2020}%
  \BibitemOpen
  \bibfield  {author} {\bibinfo {author} {\bibfnamefont {L.}~\bibnamefont
  {Farinacci}}, \bibinfo {author} {\bibfnamefont {G.}~\bibnamefont {Ahmadi}},
  \bibinfo {author} {\bibfnamefont {M.}~\bibnamefont {Ruby}}, \bibinfo {author}
  {\bibfnamefont {G.}~\bibnamefont {Reecht}}, \bibinfo {author} {\bibfnamefont
  {B.~W.}\ \bibnamefont {Heinrich}}, \bibinfo {author} {\bibfnamefont
  {C.}~\bibnamefont {Czekelius}}, \bibinfo {author} {\bibfnamefont
  {F.}~\bibnamefont {von Oppen}},\ and\ \bibinfo {author} {\bibfnamefont
  {K.~J.}\ \bibnamefont {Franke}},\ }\href
  {https://doi.org/10.1103/PhysRevLett.125.256805} {\bibfield  {journal}
  {\bibinfo  {journal} {Phys. Rev. Lett.}\ }\textbf {\bibinfo {volume} {125}},\
  \bibinfo {pages} {256805} (\bibinfo {year} {2020})}\BibitemShut {NoStop}%
\bibitem [{\citenamefont {Rubio-Verd\'u}\ \emph {et~al.}(2021)\citenamefont
  {Rubio-Verd\'u}, \citenamefont {Zald\'{\i}var}, \citenamefont {Zitko},\ and\
  \citenamefont {Pascual}}]{Verdu2021}%
  \BibitemOpen
  \bibfield  {author} {\bibinfo {author} {\bibfnamefont {C.}~\bibnamefont
  {Rubio-Verd\'u}}, \bibinfo {author} {\bibfnamefont {J.}~\bibnamefont
  {Zald\'{\i}var}}, \bibinfo {author} {\bibfnamefont {R.}~\bibnamefont
  {Zitko}},\ and\ \bibinfo {author} {\bibfnamefont {J.~I.}\ \bibnamefont
  {Pascual}},\ }\href {https://doi.org/10.1103/PhysRevLett.126.017001}
  {\bibfield  {journal} {\bibinfo  {journal} {Phys. Rev. Lett.}\ }\textbf
  {\bibinfo {volume} {126}},\ \bibinfo {pages} {017001} (\bibinfo {year}
  {2021})}\BibitemShut {NoStop}%
\bibitem [{\citenamefont {Kamlapure}\ \emph {et~al.}(2019)\citenamefont
  {Kamlapure}, \citenamefont {Cornils}, \citenamefont {Zitko}, \citenamefont
  {Valentyuk}, \citenamefont {Mozara}, \citenamefont {Pradhan}, \citenamefont
  {Fransson}, \citenamefont {Lichtenstein}, \citenamefont {Wiebe},\ and\
  \citenamefont {Wiesendanger}}]{Kamlapure2019}%
  \BibitemOpen
  \bibfield  {author} {\bibinfo {author} {\bibfnamefont {A.}~\bibnamefont
  {Kamlapure}}, \bibinfo {author} {\bibfnamefont {L.}~\bibnamefont {Cornils}},
  \bibinfo {author} {\bibfnamefont {R.}~\bibnamefont {Zitko}}, \bibinfo
  {author} {\bibfnamefont {M.}~\bibnamefont {Valentyuk}}, \bibinfo {author}
  {\bibfnamefont {R.}~\bibnamefont {Mozara}}, \bibinfo {author} {\bibfnamefont
  {S.}~\bibnamefont {Pradhan}}, \bibinfo {author} {\bibfnamefont
  {J.}~\bibnamefont {Fransson}}, \bibinfo {author} {\bibfnamefont {A.~I.}\
  \bibnamefont {Lichtenstein}}, \bibinfo {author} {\bibfnamefont
  {J.}~\bibnamefont {Wiebe}},\ and\ \bibinfo {author} {\bibfnamefont
  {R.}~\bibnamefont {Wiesendanger}},\ }\href@noop {} {} (\bibinfo {year}
  {2019}),\ \Eprint {https://arxiv.org/abs/1911.03794} {arXiv:1911.03794}
  \BibitemShut {NoStop}%
\bibitem [{\citenamefont {Odobesko}\ \emph {et~al.}(2020)\citenamefont
  {Odobesko}, \citenamefont {Di~Sante}, \citenamefont {Kowalski}, \citenamefont
  {Wilfert}, \citenamefont {Friedrich}, \citenamefont {Thomale}, \citenamefont
  {Sangiovanni},\ and\ \citenamefont {Bode}}]{Odobesko2020}%
  \BibitemOpen
  \bibfield  {author} {\bibinfo {author} {\bibfnamefont {A.}~\bibnamefont
  {Odobesko}}, \bibinfo {author} {\bibfnamefont {D.}~\bibnamefont {Di~Sante}},
  \bibinfo {author} {\bibfnamefont {A.}~\bibnamefont {Kowalski}}, \bibinfo
  {author} {\bibfnamefont {S.}~\bibnamefont {Wilfert}}, \bibinfo {author}
  {\bibfnamefont {F.}~\bibnamefont {Friedrich}}, \bibinfo {author}
  {\bibfnamefont {R.}~\bibnamefont {Thomale}}, \bibinfo {author} {\bibfnamefont
  {G.}~\bibnamefont {Sangiovanni}},\ and\ \bibinfo {author} {\bibfnamefont
  {M.}~\bibnamefont {Bode}},\ }\href
  {https://doi.org/10.1103/PhysRevB.102.174504} {\bibfield  {journal} {\bibinfo
   {journal} {Phys. Rev. B}\ }\textbf {\bibinfo {volume} {102}},\ \bibinfo
  {pages} {174504} (\bibinfo {year} {2020})}\BibitemShut {NoStop}%
\bibitem [{\citenamefont {Hirjibehedin}\ \emph {et~al.}(2007)\citenamefont
  {Hirjibehedin}, \citenamefont {Lin}, \citenamefont {Otte}, \citenamefont
  {Ternes}, \citenamefont {Lutz}, \citenamefont {Jones},\ and\ \citenamefont
  {Heinrich}}]{Hirjibehedin2007}%
  \BibitemOpen
  \bibfield  {author} {\bibinfo {author} {\bibfnamefont {C.~F.}\ \bibnamefont
  {Hirjibehedin}}, \bibinfo {author} {\bibfnamefont {C.-Y.}\ \bibnamefont
  {Lin}}, \bibinfo {author} {\bibfnamefont {A.~F.}\ \bibnamefont {Otte}},
  \bibinfo {author} {\bibfnamefont {M.}~\bibnamefont {Ternes}}, \bibinfo
  {author} {\bibfnamefont {C.~P.}\ \bibnamefont {Lutz}}, \bibinfo {author}
  {\bibfnamefont {B.~A.}\ \bibnamefont {Jones}},\ and\ \bibinfo {author}
  {\bibfnamefont {A.~J.}\ \bibnamefont {Heinrich}},\ }\href
  {https://doi.org/10.1126/science.1146110} {\bibfield  {journal} {\bibinfo
  {journal} {Science}\ }\textbf {\bibinfo {volume} {317}},\ \bibinfo {pages}
  {1199} (\bibinfo {year} {2007})}\BibitemShut {NoStop}%
\bibitem [{\citenamefont {Tsukahara}\ \emph {et~al.}(2009)\citenamefont
  {Tsukahara}, \citenamefont {Noto}, \citenamefont {Ohara}, \citenamefont
  {Shiraki}, \citenamefont {Takagi}, \citenamefont {Takata}, \citenamefont
  {Miyawaki}, \citenamefont {Taguchi}, \citenamefont {Chainani}, \citenamefont
  {Shin},\ and\ \citenamefont {Kawai}}]{Tsukuhara2009}%
  \BibitemOpen
  \bibfield  {author} {\bibinfo {author} {\bibfnamefont {N.}~\bibnamefont
  {Tsukahara}}, \bibinfo {author} {\bibfnamefont {K.-i.}\ \bibnamefont {Noto}},
  \bibinfo {author} {\bibfnamefont {M.}~\bibnamefont {Ohara}}, \bibinfo
  {author} {\bibfnamefont {S.}~\bibnamefont {Shiraki}}, \bibinfo {author}
  {\bibfnamefont {N.}~\bibnamefont {Takagi}}, \bibinfo {author} {\bibfnamefont
  {Y.}~\bibnamefont {Takata}}, \bibinfo {author} {\bibfnamefont
  {J.}~\bibnamefont {Miyawaki}}, \bibinfo {author} {\bibfnamefont
  {M.}~\bibnamefont {Taguchi}}, \bibinfo {author} {\bibfnamefont
  {A.}~\bibnamefont {Chainani}}, \bibinfo {author} {\bibfnamefont
  {S.}~\bibnamefont {Shin}},\ and\ \bibinfo {author} {\bibfnamefont
  {M.}~\bibnamefont {Kawai}},\ }\href
  {https://doi.org/10.1103/PhysRevLett.102.167203} {\bibfield  {journal}
  {\bibinfo  {journal} {Phys. Rev. Lett.}\ }\textbf {\bibinfo {volume} {102}},\
  \bibinfo {pages} {167203} (\bibinfo {year} {2009})}\BibitemShut {NoStop}%
\bibitem [{\citenamefont {Heinrich}\ \emph {et~al.}(2013)\citenamefont
  {Heinrich}, \citenamefont {Braun}, \citenamefont {Pascual},\ and\
  \citenamefont {Franke}}]{Heinrich2013b}%
  \BibitemOpen
  \bibfield  {author} {\bibinfo {author} {\bibfnamefont {B.~W.}\ \bibnamefont
  {Heinrich}}, \bibinfo {author} {\bibfnamefont {L.}~\bibnamefont {Braun}},
  \bibinfo {author} {\bibfnamefont {J.~I.}\ \bibnamefont {Pascual}},\ and\
  \bibinfo {author} {\bibfnamefont {K.~J.}\ \bibnamefont {Franke}},\ }\href
  {https://doi.org/10.1038/nphys2794} {\bibfield  {journal} {\bibinfo
  {journal} {Nat. Phys.}\ }\textbf {\bibinfo {volume} {9}},\ \bibinfo {pages}
  {765} (\bibinfo {year} {2013})}\BibitemShut {NoStop}%
\bibitem [{\citenamefont {Kezilebieke}\ \emph {et~al.}(2019)\citenamefont
  {Kezilebieke}, \citenamefont {Zitko}, \citenamefont {Dvorak}, \citenamefont
  {Ojanen},\ and\ \citenamefont {Liljeroth}}]{Kezilebieke2019}%
  \BibitemOpen
  \bibfield  {author} {\bibinfo {author} {\bibfnamefont {S.}~\bibnamefont
  {Kezilebieke}}, \bibinfo {author} {\bibfnamefont {R.}~\bibnamefont {Zitko}},
  \bibinfo {author} {\bibfnamefont {M.}~\bibnamefont {Dvorak}}, \bibinfo
  {author} {\bibfnamefont {T.}~\bibnamefont {Ojanen}},\ and\ \bibinfo {author}
  {\bibfnamefont {P.}~\bibnamefont {Liljeroth}},\ }\href
  {https://doi.org/10.1021/acs.nanolett.9b01583} {\bibfield  {journal}
  {\bibinfo  {journal} {Nano Lett.}\ }\textbf {\bibinfo {volume} {19}},\
  \bibinfo {pages} {4614} (\bibinfo {year} {2019})}\BibitemShut {NoStop}%
\bibitem [{\citenamefont {Zitko}\ \emph {et~al.}(2011)\citenamefont {Zitko},
  \citenamefont {Bodensiek},\ and\ \citenamefont {Pruschke}}]{Zitko2011}%
  \BibitemOpen
  \bibfield  {author} {\bibinfo {author} {\bibfnamefont {R.}~\bibnamefont
  {Zitko}}, \bibinfo {author} {\bibfnamefont {O.}~\bibnamefont {Bodensiek}},\
  and\ \bibinfo {author} {\bibfnamefont {T.}~\bibnamefont {Pruschke}},\ }\href
  {http://dx.doi.org/10.1103/PhysRevB.83.054512} {\bibfield  {journal}
  {\bibinfo  {journal} {Phys. Rev. B}\ }\textbf {\bibinfo {volume} {83}}
  (\bibinfo {year} {2011})}\BibitemShut {NoStop}%
\bibitem [{\citenamefont {Yao}\ \emph {et~al.}(2014)\citenamefont {Yao},
  \citenamefont {Moca}, \citenamefont {Weymann}, \citenamefont {Sau},
  \citenamefont {Lukin}, \citenamefont {Demler},\ and\ \citenamefont
  {Zar\'and}}]{Yao2014}%
  \BibitemOpen
  \bibfield  {author} {\bibinfo {author} {\bibfnamefont {N.~Y.}\ \bibnamefont
  {Yao}}, \bibinfo {author} {\bibfnamefont {C.~P.}\ \bibnamefont {Moca}},
  \bibinfo {author} {\bibfnamefont {I.}~\bibnamefont {Weymann}}, \bibinfo
  {author} {\bibfnamefont {J.~D.}\ \bibnamefont {Sau}}, \bibinfo {author}
  {\bibfnamefont {M.~D.}\ \bibnamefont {Lukin}}, \bibinfo {author}
  {\bibfnamefont {E.~A.}\ \bibnamefont {Demler}},\ and\ \bibinfo {author}
  {\bibfnamefont {G.}~\bibnamefont {Zar\'and}},\ }\href
  {https://doi.org/10.1103/PhysRevB.90.241108} {\bibfield  {journal} {\bibinfo
  {journal} {Phys. Rev. B}\ }\textbf {\bibinfo {volume} {90}},\ \bibinfo
  {pages} {241108} (\bibinfo {year} {2014})}\BibitemShut {NoStop}%
\bibitem [{\citenamefont {Allub}\ \emph {et~al.}(1981)\citenamefont {Allub},
  \citenamefont {Wiecko},\ and\ \citenamefont {Alascio}}]{Allub1981}%
  \BibitemOpen
  \bibfield  {author} {\bibinfo {author} {\bibfnamefont {R.}~\bibnamefont
  {Allub}}, \bibinfo {author} {\bibfnamefont {C.}~\bibnamefont {Wiecko}},\ and\
  \bibinfo {author} {\bibfnamefont {B.}~\bibnamefont {Alascio}},\ }\href
  {https://doi.org/10.1103/PhysRevB.23.1122} {\bibfield  {journal} {\bibinfo
  {journal} {Phys. Rev. B}\ }\textbf {\bibinfo {volume} {23}},\ \bibinfo
  {pages} {1122} (\bibinfo {year} {1981})}\BibitemShut {NoStop}%
\bibitem [{\citenamefont {Kir\u{s}anskas}\ \emph {et~al.}(2015)\citenamefont
  {Kir\u{s}anskas}, \citenamefont {Goldstein}, \citenamefont {Flensberg},
  \citenamefont {Glazman},\ and\ \citenamefont {Paaske}}]{Kirsanskas2015}%
  \BibitemOpen
  \bibfield  {author} {\bibinfo {author} {\bibfnamefont {G.}~\bibnamefont
  {Kir\u{s}anskas}}, \bibinfo {author} {\bibfnamefont {M.}~\bibnamefont
  {Goldstein}}, \bibinfo {author} {\bibfnamefont {K.}~\bibnamefont
  {Flensberg}}, \bibinfo {author} {\bibfnamefont {L.~I.}\ \bibnamefont
  {Glazman}},\ and\ \bibinfo {author} {\bibfnamefont {J.}~\bibnamefont
  {Paaske}},\ }\href {https://doi.org/10.1103/PhysRevB.92.235422} {\bibfield
  {journal} {\bibinfo  {journal} {Phys. Rev. B}\ }\textbf {\bibinfo {volume}
  {92}},\ \bibinfo {pages} {235422} (\bibinfo {year} {2015})}\BibitemShut
  {NoStop}%
\bibitem [{\citenamefont {Grove-Rasmussen}\ \emph {et~al.}(2018)\citenamefont
  {Grove-Rasmussen}, \citenamefont {Steffensen}, \citenamefont {Jellinggaard},
  \citenamefont {Madsen}, \citenamefont {Zitko}, \citenamefont {Paaske},\ and\
  \citenamefont {Nygard}}]{Grove2018}%
  \BibitemOpen
  \bibfield  {author} {\bibinfo {author} {\bibfnamefont {K.}~\bibnamefont
  {Grove-Rasmussen}}, \bibinfo {author} {\bibfnamefont {G.}~\bibnamefont
  {Steffensen}}, \bibinfo {author} {\bibfnamefont {A.}~\bibnamefont
  {Jellinggaard}}, \bibinfo {author} {\bibfnamefont {M.~H.}\ \bibnamefont
  {Madsen}}, \bibinfo {author} {\bibfnamefont {R.}~\bibnamefont {Zitko}},
  \bibinfo {author} {\bibfnamefont {J.}~\bibnamefont {Paaske}},\ and\ \bibinfo
  {author} {\bibfnamefont {J.}~\bibnamefont {Nygard}},\ }\href
  {http://dx.doi.org/10.1038/s41467-018-04683-x} {\bibfield  {journal}
  {\bibinfo  {journal} {Nat. Commun.}\ }\textbf {\bibinfo {volume} {9}}
  (\bibinfo {year} {2018})}\BibitemShut {NoStop}%
\bibitem [{\citenamefont {Estrada~Saldana}\ \emph {et~al.}(2020)\citenamefont
  {Estrada~Saldana}, \citenamefont {Vekris}, \citenamefont {Zitko},
  \citenamefont {Steffensen}, \citenamefont {Krogstrup}, \citenamefont
  {Paaske}, \citenamefont {Grove-Rasmussen},\ and\ \citenamefont
  {Nyg{\aa}rd}}]{Saldana2020}%
  \BibitemOpen
  \bibfield  {author} {\bibinfo {author} {\bibfnamefont {J.~C.}\ \bibnamefont
  {Estrada~Saldana}}, \bibinfo {author} {\bibfnamefont {A.}~\bibnamefont
  {Vekris}}, \bibinfo {author} {\bibfnamefont {R.}~\bibnamefont {Zitko}},
  \bibinfo {author} {\bibfnamefont {G.}~\bibnamefont {Steffensen}}, \bibinfo
  {author} {\bibfnamefont {P.}~\bibnamefont {Krogstrup}}, \bibinfo {author}
  {\bibfnamefont {J.}~\bibnamefont {Paaske}}, \bibinfo {author} {\bibfnamefont
  {K.}~\bibnamefont {Grove-Rasmussen}},\ and\ \bibinfo {author} {\bibfnamefont
  {J.}~\bibnamefont {Nyg{\aa}rd}},\ }\href@noop {} {\bibfield  {journal}
  {\bibinfo  {journal} {Phys. Rev. B}\ }\textbf {\bibinfo {volume} {102}},\
  \bibinfo {pages} {195143} (\bibinfo {year} {2020})}\BibitemShut {NoStop}%
\bibitem [{\citenamefont {von Oppen}\ and\ \citenamefont
  {Franke}(2021)}]{Oppen2021}%
  \BibitemOpen
  \bibfield  {author} {\bibinfo {author} {\bibfnamefont {F.}~\bibnamefont {von
  Oppen}}\ and\ \bibinfo {author} {\bibfnamefont {K.~J.}\ \bibnamefont
  {Franke}},\ }\href {https://doi.org/10.1103/PhysRevB.103.205424} {\bibfield
  {journal} {\bibinfo  {journal} {Phys. Rev. B}\ }\textbf {\bibinfo {volume}
  {103}},\ \bibinfo {pages} {205424} (\bibinfo {year} {2021})}\BibitemShut
  {NoStop}%
\bibitem [{\citenamefont {Sakurai}(1970)}]{Sakurai1970}%
  \BibitemOpen
  \bibfield  {author} {\bibinfo {author} {\bibfnamefont {A.}~\bibnamefont
  {Sakurai}},\ }\href {https://doi.org/10.1143/PTP.44.1472} {\bibfield
  {journal} {\bibinfo  {journal} {Prog. Theor. Phys.}\ }\textbf {\bibinfo
  {volume} {44}},\ \bibinfo {pages} {1472} (\bibinfo {year}
  {1970})}\BibitemShut {NoStop}%
\bibitem [{\citenamefont {Otte}\ \emph {et~al.}(2008)\citenamefont {Otte},
  \citenamefont {Ternes}, \citenamefont {von Bergmann}, \citenamefont {Loth},
  \citenamefont {Brune}, \citenamefont {Lutz}, \citenamefont {Hirjibehedin},\
  and\ \citenamefont {Heinrich}}]{Otte2008}%
  \BibitemOpen
  \bibfield  {author} {\bibinfo {author} {\bibfnamefont {A.~F.}\ \bibnamefont
  {Otte}}, \bibinfo {author} {\bibfnamefont {M.}~\bibnamefont {Ternes}},
  \bibinfo {author} {\bibfnamefont {K.}~\bibnamefont {von Bergmann}}, \bibinfo
  {author} {\bibfnamefont {S.}~\bibnamefont {Loth}}, \bibinfo {author}
  {\bibfnamefont {H.}~\bibnamefont {Brune}}, \bibinfo {author} {\bibfnamefont
  {C.~P.}\ \bibnamefont {Lutz}}, \bibinfo {author} {\bibfnamefont {C.~F.}\
  \bibnamefont {Hirjibehedin}},\ and\ \bibinfo {author} {\bibfnamefont {A.~J.}\
  \bibnamefont {Heinrich}},\ }\href {https://doi.org/10.1038/nphys1072}
  {\bibfield  {journal} {\bibinfo  {journal} {Nat. Phys.}\ }\textbf {\bibinfo
  {volume} {4}},\ \bibinfo {pages} {847} (\bibinfo {year} {2008})}\BibitemShut
  {NoStop}%
\bibitem [{Note1()}]{Note1}%
  \BibitemOpen
  \bibinfo {note} {\protect \citet {Yao2014} assumed a particle-hole symmetric
  substrate which gives rise to a level crossing between the local-singlet and
  molecular-singlet states. As we do not assume this additional symmetry, we
  observe the more generic avoided crossing.}\BibitemShut {Stop}%
\bibitem [{Note2()}]{Note2}%
  \BibitemOpen
  \bibinfo {note} {We note that the energy balance inverts, when $K_\perp
  \lesssim \abs {D}S^4$. Corresponding phase diagrams are included and
  discussed in App.\ \ref {app:higher_spin_dimer_iso_dmi}.}\BibitemShut {Stop}%
\bibitem [{\citenamefont {Mishra}\ \emph {et~al.}(2021)\citenamefont {Mishra},
  \citenamefont {Simon}, \citenamefont {Hyart},\ and\ \citenamefont
  {Trif}}]{Mishra2021}%
  \BibitemOpen
  \bibfield  {author} {\bibinfo {author} {\bibfnamefont {A.}~\bibnamefont
  {Mishra}}, \bibinfo {author} {\bibfnamefont {P.}~\bibnamefont {Simon}},
  \bibinfo {author} {\bibfnamefont {T.}~\bibnamefont {Hyart}},\ and\ \bibinfo
  {author} {\bibfnamefont {M.}~\bibnamefont {Trif}},\ }\href
  {https://doi.org/10.1103/PRXQuantum.2.040347} {\bibfield  {journal} {\bibinfo
   {journal} {PRX Quantum}\ }\textbf {\bibinfo {volume} {2}},\ \bibinfo {pages}
  {040347} (\bibinfo {year} {2021})}\BibitemShut {NoStop}%
\bibitem [{\citenamefont {Pavesic}\ and\ \citenamefont
  {Zitko}(2022)}]{Pavesic2021}%
  \BibitemOpen
  \bibfield  {author} {\bibinfo {author} {\bibfnamefont {L.}~\bibnamefont
  {Pavesic}}\ and\ \bibinfo {author} {\bibfnamefont {R.}~\bibnamefont
  {Zitko}},\ }\href {https://doi.org/10.1103/PhysRevB.105.075129} {\bibfield
  {journal} {\bibinfo  {journal} {Phys. Rev. B}\ }\textbf {\bibinfo {volume}
  {105}},\ \bibinfo {pages} {075129} (\bibinfo {year} {2022})}\BibitemShut
  {NoStop}%
\end{thebibliography}

%

\end{document}